\documentstyle[12pt,epsfig]{article}
\newcommand{\be}{\begin{eqnarray}}
\newcommand{\ee}{\end{eqnarray}}
\newcommand{\del}{\partial}

\newcommand{\Pslash}{P\hspace{-.5em}/\hspace{.15em}}
\newcommand{\pslash}{p\hspace{-.5em}/\hspace{.15em}}

\newcommand{\vect}[1]{{\mbox{\boldmath $#1$}}}
\begin{document}
%
%
\rightline{UNITUE--THEP--6/98}
\rightline{nucl-th/9805054}
\begin{flushright}
\framebox{Draft : \today}
\end{flushright}
%
\begin{center}
\begin{large}
{\bf Octet and Decuplet Baryons \\
in a Covariant and Confining Diquark-Quark Model$^\dagger$\\}
\end{large}
\vspace{1cm}
{\bf M. Oettel$^*$, G. Hellstern$^\$ $,  
R. Alkofer$^\#$ and H. Reinhardt\\}
\vspace{0.2cm}
{\em Institute for Theoretical Physics,\\
T\"ubingen University,\\
Auf der Morgenstelle 14,\\
D--72076 T\"ubingen, Germany} \\ [0.5cm]
\end{center}
\vspace{1cm}
\begin{abstract}
\noindent
The baryon octet and decuplet masses and Bethe-Salpeter vertex and 
wave functions
are calculated in the ladder approximation to the quark exchange 
between a scalar or axialvector diquark and a constituent quark.
These functions reflecting full Lorentz covariance are given
in terms of an expansion in Gegenbauer polynomials.
In the rest frame of the baryon, a complete partial wave
decomposition of the Bethe-Salpeter wave function is performed.
The confinement of quarks and diquarks is implemented via a 
parametrisation  of the corresponding propagators.
We also discuss some aspects of the momentum routing
in the ladder approximation to the
Bethe-Salpeter equation.
Numerical results for the octet and decuplet masses with 
broken flavour $SU(3)$ in the conserved isospin limit 
are presented.
\end{abstract}
\vspace{.5cm}
\vfill
\rule{5cm}{.15mm}
\\
\noindent
{\footnotesize $^\dagger$ Supported by BMBF 
under contract 06TU888, DFG under contract
We 1254/4-1 and Graduiertenkolleg ``Hadronen und Kerne''
(DFG Mu705/3).\\
$^*$ E-mail: oettel@pthp3.tphys.physik.uni-tuebingen.de\\
$^\$ $Address after June ${\rm 1}^{st}$, 1998: 
Deutsche Bank, Frankfurt.\\
$^\#$ E-mail: Reinhard.Alkofer@uni-tuebingen.de}
\newpage
%
%
%
\section{Introduction}
In a recent paper \cite{Hel97} we presented numerical results for the nucleon 
form factors in a fully Lorentz covariant model based on the idea 
that baryons may be viewed as bound states of confined constituent 
diquarks and quarks interacting via quark exchange. The confinement
of the constituents is hereby effectively parametrised.
The poles of the corresponding propagators are removed by some 
modifying multiplicative factors which, however, possess an essential 
singularity at time-like infinite momentum. The
physical picture behind this baryon model is very natural: Diquarks 
are allowed to ``decay'' into two quarks, one of them recombines
with the third quark and forms another diquark. The hope is,
that for physics relevant at intermediate momentum 
transfers, most of the complicated structure of the baryon
may be efficaciously described by assuming strong correlations
in the quark-quark channel. Thus the notion
of diquarks parametrises to some extend unknown non-perturbative physics 
within baryons. In the last years diquarks have not only been 
used in non-perturbative calculations but also in the description 
of inelastic lepton-nucleon scattering, see refs. \cite{Ans93,Ans97}
where various applications of diquarks are extensively 
discussed. Very recently diquark masses have been estimated from 
lattice measurements \cite{Hes98}.

The main purpose of our investigations is to formulate 
a baryon model applicable to the intermediate energy region.
This is mainly motivated with the 
advent of a new generation of 
continuous beam facilities
like CEBAF at TJNAF, MAMI, ELSA, COSY 
etc. which are designed to explore
an intermediate region lying between the non-perturbative
low-energy and the perturbative high-energy regime of QCD.
These facilities explore various hadron observables
to a very high precision.
The different existing 
hadron models are to be judged 
by their ability to predict and to explain these
observables in the near future.
While there exist many models capable to 
describe pion properties
which  are strongly 
dictated by chiral symmetry, the detailed structure
of all other mesons are still unclear.
Although quirte a few baryon models, see e.g. 
\cite{Sky61,Kar68,Fai68,Fey71,Cho75,Has78,Rho94},
were developed in the last forty years, a unified 
description of baryons within a covariant field theoretical 
approach, and with quarks and gluons as the fundamental
degrees of freedom,   
is still missing. Besides being covariant, such 
a description should include chiral symmetry in its
spontaneously broken phase and confinement.
A covariant model with the correct symmetry pattern
but without confinement is the Nambu--Jona-Lasinio (NJL) model 
\cite{Nam61,Nam61a}
where quarks as fundamental fermion fields
interact locally, for recent reviews see
\cite{Hat94,Ebe94,Alk95}.
There, baryons appear  either as non-topological
solitons \cite{Rei88,Alk96,Chr96} or as bound states
of a quark and a diquark \cite{Rei90,Vog91,Buc92,Buc95,Hua94,Han95,Ish93}.
A hybrid formalism combining the soliton with the 
diquark picture has been developed in ref. \cite{Zue97},
where it turned out, that the soliton background contributes 
as much to the total binding energy of the nucleon as
the direct coupling between two quarks and between quark and diquark.
In the Global Colour Model, a non-local extension
of the NJL model, there also exist preliminary
studies of nucleons as diquark-quark bound states, see e.g.
\cite{Cah92} and references therein.

In this paper we extend the investigations reported in
ref. \cite{Hel97}. There we solved the Bethe-Salpeter 
equation for nucleons in ladder approximation but 
restricted ourselves to scalar diquarks. Employing
the solution of the Bethe-Salpeter equation, i.e. the
nucleon vertex functions, we calculated
various form factors like the 
electromagnetic, the weak and the pionic
form factor of the nucleon. Despite the fact
that this work is 
a promising starting point for further 
investigations, some results, especially for the magnetic
moments, signalled that the axialvector diquark channel
is necessary for a realistic description of baryons
as bound states of quarks and diquarks.
Here we therefore include the 
axialvector diquark 
channel into the nucleon Bethe-Salpeter equation.
We also present results for baryons with spin
$3/2$. 

The paper is organised as follows: In the next section the covariant
and confining diquark-quark model, as defined in ref.\cite{Hel97}, is 
briefly reviewed. By modifying the propagators of quarks
and diquarks an effective modelling of confinement
enters our model. In section \ref{BSEchap}
the Bethe-Salpeter
equations for spin-1/2 and spin-3/2 baryons, determining their masses
and wave functions, are discussed. With an appropriate three
spinor basis we then construct  covariant
ans{\"a}tze for the wave functions which are  suitable for the numerical 
solution of the homogeneous integral equations.
Section \ref{Rootchap} is devoted to a discussion of the
subtleties concerning the momentum routing in  
the Bethe-Salpeter
equation, associated with the ladder approximation.
The numerical method is shortly described in section
\ref{Numchap}, and our results for the baryon masses and wave functions
are described  in section \ref{Reschap}. In the last section we finally
conclude and give an outlook. Some technical issues are deferred 
to three appendices.
  
\section{The Covariant and Confining \\
Diquark-Quark Model}
Here we briefly recapitulate the definition 
of the covariant and confining diquark-quark model as given 
in ref. \cite{Hel97}. Such a model, although without confinement
has been also considered in
\cite{Mey94,Kus97}.
Since the solution of a 
rigorous relativistic three body equation, a Faddeev equation \cite{Fad61},
is still missing in field theory
we follow the path of using  diquarks as effective
degrees of freedom within baryons.  
They serve
as an efficient tool to parametrise some of the
unknown non-perturbative features of the baryon wave function. 
As stated in the introduction such approaches have been
successful in the past to describe
baryons in NJL-type models \cite{Rei90,Vog91,Buc92,Buc95,Hua94,Han95,Ish93}.
A three-dimensional reduction of the fully covariant Bethe-Salpeter
model discussed here, can be found in ref. \cite{Kei96},
where the author solved the bound state equations in the 
Salpeter approximation which ignores all retardation effects.

Having in mind the  diquark-quark description arising 
within 
the hadronized NJL model \cite{Rei90}, some basic 
structures are fixed: 
To build up a colourless baryon out of a diquark
and a quark being in the fundamental 
representation of the colour group $\rm{SU(3)_C}$,
diquarks 
necessarily live in the colour anti-triplet channel.
Furthermore, in order to fulfil the Pauli principle,
scalar diquarks
couple via the antisymmetric generators of the flavour group,
$t_{\mathcal A}^a=\{\rho^{a=1..3}\}$, axialvector diquarks
via the symmetric generators
$t_{\mathcal S}^a=\{\rho^{a=4..9}\}$, respectively \cite{Rei90}.
Our conventions for these flavour matrices $\rho^a$ are given in  
appendix \ref{flavoureq}.  
In the following, however, 
we deviate from the NJL model as described in \cite{Alk95}.
Quarks and diquarks are treated as elementary
but confined particles, see below, whose interaction, quark
exchange, gives rise to quark-diquark correlations strong
enough to bind these fields to a baryon.

This can be formalised with the Lagrangian
\be 
\label{L1} 
\cal L & = &  \bar q_A(x) (i\gamma^\mu\partial_\mu-{\bf m}_q)
f(-\partial^2/{\bf m}_q^2)
q_A(x)
\nonumber \\ 
& &{} 
+ \Delta^\dagger_A (x)(-\partial_\mu\partial^\mu - {\bf m}_{0^+}^2)
f(-\partial^2/{\bf m}_{0^+}^2) 
   \Delta _A (x)
\nonumber \\ 
& &{}
 - \frac{1}{4}F^\dagger_{\mu\nu}(x)f(-\partial^2/{\bf m}_{1^+}^2)F^{\mu\nu}(x) + 
\frac{1}{2}{\bf m}_{1^+}^2\Delta^\dagger _{Aa\mu }(x)f(-\partial^2/{\bf m}_{1^+}^2)
\Delta^\mu _{Aa} (x) 
\nonumber \\ 
& &{} + \epsilon^{ABC} \left(
 g_s q_C^{T}(x) C i\gamma^5 t_{\mathcal A}^a q_B(x) \Delta_{aA}^{*}(x) 
+g_s^{*} \Delta_{aA}(x)\bar q_B(x) i\gamma^5 C  t_{\mathcal A}^{\dagger a} 
\bar q_C^T(x) \right)
\nonumber \\ 
& &{} + \epsilon^{ABC} \left(
 g_a q_C^{T}(x) C i\gamma^{\mu} t_{\mathcal S}^a q_B (x)
 \Delta^{*}_{Aa\mu}(x)
-g_a^* \Delta_{Aa\mu}(x)\bar{q}_B (x) t_{\mathcal S}^{\dagger a} 
 i\gamma^{\mu} C \bar{q}_C^{T} (x)\right).
\ee 
The quark field is denoted by $q(x)$, the scalar diquark field
by $\Delta(x)$ and the axialvector diquark field 
by $\Delta^\mu(x)$. Their masses are given by the matrices
in flavour space  ${\bf m}_q$,${\bf m}_{0^+}$ and
${\bf m}_{1^+}$, respectively. For unbroken flavour symmetry
they reduce to ${\bf m}_q$=diag$_3(m_q)$, ${\bf m}_{0^+}$=diag$_3(m_{0^+})$
and ${\bf m}_{1^+}$=diag$_6(m_{1^+})$.
In eq. (\ref{L1}), capital subscripts denote colour quantum numbers.
In the kinetic part of the $1^+$ diquark
the non-abelian field strength tensor 
$F^{\mu \nu} = \del^\mu \Delta^\nu - \del^\nu \Delta^\mu + [\Delta^\mu,
\Delta^\nu]$ appears ($\Delta_\mu=\Delta_{a\mu}t_{\mathcal S}^a$). 
The associated self-interactions, however, will not be taken into account.
The coupling strengths, of Yukawa type,
between  two quarks and the scalar or  axialvector
diquarks are given by $g_s$ or  $g_a$, respectively. 
In order to take an extended diquark into account, the point-like
couplings will be supplemented with momentum dependent factors to be 
defined later.
 
An essential ingredient entering our model is the
effective parametrisation of confinement. This is realised
by modifying the kinetic terms of the constituents
in the Lagrangian (\ref{L1}). Going to Euclidean\footnote{
We use an Euclidean space formulation
with $\{\gamma_\mu,\gamma_\nu\} = 2 \delta_{\mu \nu}$, $\gamma_\mu^\dagger
= \gamma_\mu$ and $pq = \sum_{\mu=1}^{4} p_\mu q_\mu $.}  momentum space and using  
\be
f^{-1}(x) = 1- e^{-d(1+x)}.
\label{conff}
\ee 
the model 
quark and diquark propagators, being diagonal in colour and flavour space,
are given by
\be
S(p) &=&  \frac{i{p  \!\!\! /}-m_{q}}{p^2+m_{q}^2}
\left( 1-e^{- d(p^2+m_{q}^2)/{m_{q}^2}} \right) ,
\label{S} \\
D(p) &=&   -\frac{1}{p^2+m_{0^+}^2}
\left( 1-e^{-d(p^2+m_{0^+}^2)/{m_{0^+}^2}} \right),
\label{Ds} \\
D^{\mu\nu} (p) &=& -\frac{
(\delta^{\mu\nu} + p^{\mu}p^{\nu}/{m_{1^+}^2})}
{p^2+m_{1^+}^2}
\left( 1-e^{-d(p^2+m_{1^+}^2)/{m_{1^+}^2}} \right).
\label{Da} 
\ee
Due to the numerators, the mass poles of the  propagators
are effectively screened\footnote{With $f\equiv 1$ we refer
to the propagators as tree level propagators.}. The strength of the
screening is described  by the parameter $d$.
Whereas the analytic behaviour of the quark propagator
(\ref{S}) is similar to the ones obtained from Dyson-Schwinger
studies of QCD (details about this approach can be found
in ref.\cite{Rob94}), the justification for using
confined  diquarks is somewhat more 
involved. As stated above, diquarks are not colour
singlets and should therefore be confined by the fundamental
interaction of QCD. Nevertheless
most diquark models \cite{Cah87,Tho90,Wei93} which  describe  
them as bound states of two quarks predict diquarks to be
observable particles. Recent studies, however, which 
investigated the system of the quark Dyson-Schwinger equation
and the diquark Bethe-Salpeter equation  beyond the 
usually  employed rainbow-ladder approximation,
were able to explain why diquarks do not appear in the
observable particle spectrum. This mechanism works in the
Munczek-Nemirovsky model \cite{Mun83}, as has been shown  
in ref. \cite{Ben96}, as well as  in an extended 
NJL model \cite{Hel97a}. Since both models assume a simplified
but quite 
different quark-quark interaction, one may conjecture that the
same also holds true implementing a realistic interaction.
For these reasons we use the notion of confined diquarks.
In our approach we therefore use the propagators defined
in eqs. (\ref{Ds}, \ref{Da}) which do not allow 
observable diquarks. In the following it will be seen,
that working with confined quarks and diquarks 
leads to the absence of an  unphysical quark-diquark threshold for baryons.

Additionally we introduce the diagonal
approximation to the axialvector diquark propagator
by omitting the $p^\mu p^\nu/m_{1+}^2$ term,
\be \label{propdiag}
 D^{\mu\nu}(p)=-\frac{\delta^{\mu\nu}}{p^2+m_{1^+}^2}f^{-1}\left(\frac{p^2}{m_{1^+}^2}
 \right).
\ee
It is well-known that the full
Proca propagator may cause spurious ultraviolet problems.
These vanish in the special, 
unitary gauge and have no physical consequences. 
As such a treatment is, however, beyond the scope of our 
purely phenomenological investigation
we have chosen to avoid
this problem by using a diagonal  approximation to the
propagator.
Its validity is 
examined more closely in appendix \ref{Procadis}.

\section{Bethe-Salpeter Equation for Octet and Decuplet Baryons
  in the Flavour-Symmetric Case}
\label{BSEchap}
Using the Lagrangian given in eq. (\ref{L1}) we obtain
the ladder Bethe-Salpeter equation for octet and decuplet baryons,

\begin{eqnarray} \label{sc} 
 \pmatrix{\Psi_8(p;P) \cr \Psi^{\nu}_8(p;P) \cr} &=& - |g_s|^2
  \pmatrix{D(p_b) & 0 \cr 0 & D^{\nu\mu}(p_b) \cr}
  \, S(p_a) \\
  & & \times \int\frac{d^4p^{\prime}}{(2\pi)^4}
 \pmatrix{ \gamma_5 \tilde S(-q) \gamma_5 & -\sqrt{3} \frac{g_a}{g_s} \, 
 \gamma^{\mu'} \tilde S(-q) \gamma_5 \cr
 -\sqrt{3} \frac{g_a^*}{g_s^*} \, \gamma_5 \tilde S(-q) \gamma^{\mu}
 & -\frac{|g_a|^2}{|g_s|^2} \,\gamma^{\mu'} \tilde S(-q) \gamma^{\mu} \cr} 
 \, \pmatrix{\Psi_8(p';P)
  \cr \Psi^{\mu'}_8(p';P) \cr} \nonumber \\
 & &\nonumber\\
\Psi^{\nu \rho}_{10}(p;P)& =&
- 2 |g_{a}|^{2} S(p_{a}) D^{\nu \mu}(p_{b})  
\int\frac{d^4p^{\prime}}{(2\pi)^4}
\gamma^{\lambda} \tilde S(-q) \gamma^{\mu} 
\Psi^{\lambda \rho}_{10}(p^{\prime};P).  
\label{deltaBSE}
\end{eqnarray}
In this formulation, the Bethe-Salpeter equation involves
the matrix-valued Bethe-Salpeter wave functions of octet baryons
$\Psi_8(p;P)$ and $\Psi^\nu_8(p;P)$ and decuplet baryons
$\Psi^{\lambda \nu}_{10}(p;P)$, respectively,
which are  projected on positive parity and spin 1/2
or  spin 3/2 (see the following subsection for their construction).
Their flavour part
is given by pure octet and decuplet states in $SU(3)_{flavour}$. 
They depend on the total momentum $P$ of the bound state
and on the relative momentum $p'$ or $p$ bet\-ween the two constituents.
Mathematically, the Bethe-Salpeter equations are equivalent to
coupled 
homogeneous integral equations.
The numerical  method for its solution is  presented
in sect. \ref{Numchap}, see also \cite{Oet98}.

Although the Lagrangian (\ref{L1}) describes 
a renormalisable
diquark-quark theory, at leat at one-loop level, 
the Bethe-Salpeter equations in ladder 
approximation are formally divergent in the ultraviolet.
The divergence, of course, should then be cured by the wave 
functions. However,
to crudely take into account the extended nature of diquarks
we work, as in ref. \cite{Kus97,Hel97}, 
with a finite interaction in momentum space
and modify the propagator of the exchanged quark according to
\be
S(q) \rightarrow \tilde S(q) =
S(q)\left( \frac{\Lambda^2}{q^2+\Lambda^2}\right).
\label{dpff}
\ee
This corresponds to a monopole-type form factor.
As a consequence, this also removes all formal ultraviolet divergencies.
Note that we have absorbed the charge conjugation 
matrix $C$ appearing in the Lagrangian (\ref{L1})
using the identity $C^{-1}\tilde S^T(q) C= \tilde S(-q)$.

Before actually solving the integral equations it is 
appropriate to find a suitable basis for the
wave functions appearing in eqs. (\ref{sc},\ref{deltaBSE}).
\pagebreak
\subsection{Relativistic Three Quark States and their Wave Functions}
\label{Ansatzchap}
When constructing a baryon out of three
quarks, its wave function is 
formally described by a spinor of rank three, 
$\psi_{\alpha \beta \gamma}$ $(\alpha,\beta,\gamma =1 \ldots 4)$.
For octet baryons, a convenient  basis for this 
multi-spinor
can be found by expanding the direct product of the spinors
describing quarks of flavour $b$ and $c$ with spinor indices $\beta$ and  
$\gamma$ into the complete set of Dirac matrices
and taking the direct product with a spinor basis
of quark $a$ \cite{Hen75}:
\be \label{3qstate}
 \psi^8_{\alpha\beta\gamma}=(\Gamma u)_\alpha(\phi C\gamma_5)_{\beta\gamma}.
\ee 
$C$ denotes charge conjugation and $u=\left\{\pmatrix{\chi^+\cr 0\cr},
\pmatrix{\chi^-\cr 0\cr}\right\}$ is a basis of
positive energy Dirac spinors in the rest frame describing fermions with
spin up ($+$) and spin down ($-$). 
In this representation $\Gamma$ and $\phi$
are Dirac matrices to be expanded in the complete set
$\{{\bf 1}, \gamma_5, \gamma^\mu,\gamma_5 \gamma^\mu,\sigma^{\mu \nu}\}$.

Under a Lorentz transformation $S(\Lambda)$, this wave function
transforms according to
\be
 \psi^8_{\alpha\beta\gamma}=
(S \Gamma(\Lambda^{-1}P, \Lambda^{-1}p) S^{-1})_{\alpha \alpha'} 
(S u)_{\alpha'}(S \phi S^{-1} C\gamma_5)_{\beta\gamma}.
\ee
The parity transformation is given by
\be
 \psi^8_{\alpha\beta\gamma}
=\pi (\gamma^4 \Gamma(\tilde \gamma^\mu, \tilde P, \tilde p) u)_\alpha
(\gamma^4 \phi(\tilde \gamma^\mu, \tilde P, \tilde p) \gamma^4
 C\gamma_5)_{\beta\gamma},
\ee
where $\pi$ is the intrinsic parity of the baryon and 
$\tilde P =(-{\bf P}, P^4)$, 
$\tilde p =(-{\bf p}, p^4)$,  
$\tilde \gamma^\mu =(-\vect{\gamma}, \gamma^4)$.

To ensure total spin 1/2 of the wave function, the free Lorentz
indices in (\ref{3qstate}) are to be contracted with the independent momenta
involved and each covariant will be multiplied by a scalar function.

In the diquark-quark model, only the two choices for 
$\phi=\{{\bf 1}, \gamma^\mu\gamma_5\}$ are taken into account
which describe the scalar and axialvector diquark. 
The covariants
which lead to positive parity of spin-1/2 octet
states, 
$(\Gamma u)_\alpha$,
can be grouped according to scalar and axialvector diquark states
with the spinor indices $\beta$ and $\gamma$, see table 
\ref{covtable}, second row. As described above, the free Lorentz indices in these
covariants are contracted with the momenta $P$ and $p$ and  are then
to be multiplied with
scalar functions. No further symmetrisation
of the wave function is necessary as this will be provided by the quark 
exchange.

\begin{table}[t]
\vspace{0.2cm}
\centering
\begin{tabular}{|l|l|l|} 
\hline
 & Scalar  Diquark - $(C\gamma_5)_{\beta\gamma}$  
 & Axialvector Diquark - $(\gamma_\mu \gamma_5 C\gamma_5)_{\beta\gamma}$ \\[2mm] 
\hline
 Octet    & $(\Gamma_i^S u)_\alpha$ & $(\Gamma_i^{A \mu} u)_\alpha $ \\[2mm]
\hline 
Decuplet &                         & $(\Gamma_i^S u^\mu)_\alpha $ \\
          &                         & $(\gamma_5\, \Gamma^{A \mu}_i\, 
                                      [p^\nu_T u_\nu])_\alpha$\\[2mm] 
\hline
\hline
$\Gamma_i^S \in $ & \multicolumn{2}{|l|}{
$ \{ {\bf 1}, \,\Pslash, \, 
\pslash, \, 
P_\mu\sigma^{\mu\nu}p_\nu \} $ } 
\\[2mm]
\hline 
$\Gamma_i^{A \mu} \in $ & \multicolumn{2}{|l|}{
$ \{ P^\mu\gamma_5, \,
  p^\mu\gamma_5, \,
  \gamma^\mu \gamma_5, \,
  P^\mu\gamma_5\Pslash, \,
  P^\mu\gamma_5\pslash, $} \\
                        & \multicolumn{2}{|l|}{
$  p^\mu\gamma_5\Pslash, \,     
  p^\mu\gamma_5\pslash, \, 
  P_\nu\gamma_5\sigma^{\nu\mu}, \,
  p_\nu\gamma_5\sigma^{\nu\mu}, $} \\
                        & \multicolumn{2}{|l|}{
$ P^\mu\gamma_5P_\nu\sigma^{\nu\rho}p_\rho , \,
  p^\mu\gamma_5P_\nu\sigma^{\nu\rho}p_\rho , \,
  \gamma^\mu\gamma_5P_\nu\sigma^{\nu\rho}p_\rho \} $}
\\[2mm]
\hline
\end{tabular}         
\caption{\label{covtable}{\sf 
The Lorentz covariants leading to baryons with positive
parity. They are grouped according to their diquark content.}}
\end{table}

Using  this classification scheme, 
the octet wave function can be  denoted by
\be \label{oct1}
 \psi^8_{\alpha\beta\gamma}=\sum_{i=1}^4
(Z_i^S \Gamma_i^S u)_\alpha (C\gamma_5)_{\beta\gamma} + \sum_{i=5}^{16}
   (Z_i^A \Gamma_i^{A\mu} u)_\alpha (\gamma_\mu C)_{\beta\gamma}.
\ee
Each of the four covariants $\Gamma^i_S$ which describe the scalar diquark
part in the octet baryon wave function is multiplied by a scalar function $Z_i^S$ and 
likewise there are twelve scalar functions $Z_i^A$  multiplying the
axialvector diquark covariants $\Gamma^{A\mu}_i$.

A further reduction of this ansatz by a projection 
to positive energies is very convenient.
From the expression (\ref{oct1}) for
the octet wave function we consider only the part with spinor index $\alpha$,
multiply it
with the adjoint spinor $\bar u(P,s)$ and sum over the spins.
This 
leads to a wave function which is, by construction,
an eigenfunction of the
positive-energy
projector\footnote{With a hat we denote normalised four vectors, e.g. 
$\hat P\! \cdot \! \hat P = 1$. In the Euclidean rest frame
$P=(\vect{0},iM)$ this explicitly reads $\hat P=P/iM$. Note that
all relative momenta ($p$, $p'$) are real in Euclidean space as they are
only needed for spacelike values.} 
$\Lambda^+=(1+\hat \Pslash)$. Thus we are led
to the following relations between the wave functions of eq. (\ref{oct1})
and the corresponding functions in the Bethe-Salpeter equation (\ref{sc}):
\be
\label{eig}
\pmatrix{Z^S_{i} {\Gamma}^S_i \cr Z^A_{i} {\Gamma}^{A \mu}_i \cr}
\Lambda^+ = 
\pmatrix{S_{i} {\cal S}_i \cr A_{i} {\cal A}^{\mu}_i \cr}
=
\pmatrix{S_{i} {\cal S}_i \cr A_{i} {\cal A}^{\mu}_i \cr}
\Lambda^+
=
\pmatrix{\Psi_8(p;P) \cr \Psi^{\mu}_8(p;P) \cr}
. \label{NUCGEN}
\ee
The independent covariants ${\cal S}_i$ and ${\cal A}^\mu_i$ are required to
be eigenfunctions of $\Lambda^+$ which reduces the number of independent scalar
functions from sixteen 
to eight which are now denoted by $S_i$, (i=1,2) and $A_i$, (i=1\dots 6).
A convenient representation of these covariants
suitable for our numerical procedure and for further applications is given by
\begin{eqnarray}
 {\cal S}_i&=&\left\{ \begin{array}{l}{\cal S}_1=\Lambda^+ \\
                              {\cal S}_2=-\frac{i}{p} \pslash_T\Lambda^+ 
                              \end{array} \right. \label{scov} \\
 {\cal A}^\mu_i&=&\left\{ \begin{array}{l}
       {\cal A}^\mu_1=-\frac{i}{p}\hat P^\mu \gamma_5\pslash_T\Lambda^+ \\
       {\cal A}^\mu_2=\hat P^\mu \gamma_5\Lambda^+ \\
      {\cal A}^\mu_3=\hat p^\mu_T \gamma_5\hat \pslash_T \Lambda^+ \\
     {\cal A}^\mu_4=\frac{i}{p} p^\mu_T \gamma_5\Lambda^+ \\
      {\cal A}^\mu_5=\gamma_5\gamma^\mu_T \Lambda^+-{\cal A}^\mu_3 \\
     {\cal A}^\mu_6=
                  \frac{i}{p}\gamma_5\gamma^\mu_T \pslash_T \Lambda^+-{\cal A}^\mu_4,
                \end{array}  \label{acov} \right.
\end{eqnarray}                 
where $\gamma^\mu_T=\gamma^\mu-\hat P^\mu \hat\Pslash$.
Note that the indices have been chosen such that matrices 
with odd indices $i=\{1,3,5\}$ are eigenfunctions to $\Pslash -iM$
whereas the ones with even indices $i=\{2,4,6\}$
are eigenfunctions to $\Pslash +iM$, with eigenvalue 0 in both cases.                              

In the rest frame of the bound state, $P=({\bf 0},iM)$, 
the ansatz for the matrix valued
nucleon wave function which we used for further numerical processing reads

\be
\left( \begin{array}{c}
\Psi_8(p;P) \\
\Psi^{4}_8(p;P) \\
{\vect{\Psi}}_8(p;P)\\ 
\end{array}\right)
=
\left( \begin{array}{c}
\left( \begin{array}{cc}
{\bf 1} \, {S}_1  & 0\\
\frac{1}{p}({\vect{\sigma}}\vect{p})\, {S}_2 & 0 \\ 
\end{array}\right)  \\
\left( \begin{array}{cc}
\frac{1}{p}(\vect{\sigma}\vect{p})\, {A}_1 & 0\\ 
{\bf 1} \, {A}_2 & 0\\
\end{array}\right) \\
\left( \begin{array}{cc}
i\hat{\vect{p}}(\vect{\sigma}\hat{\vect{p}}){A}_3 + 
   (\vect{\sigma} \times \hat{\vect{p}})
   (\vect{\sigma}\hat{\vect{p}}){A}_5 & 0\\
\frac{i}{p}\vect{p}{A}_4 + \frac{1}{p}(\vect{\sigma} \times 
{\vect{p}}) {A}_6 & 0 \\
\end{array}\right) 
\end{array}\right)
.
\label{expnuc}
\ee 
\vspace{2mm}

The scalar quantities $S_i$ and $A_i$ 
depend on $p = \sqrt{p_\mu p_\mu}$ and $z= \cos \psi 
= \hat P \cdot \hat p$ for each value of the bound state mass.
As expected from the properties of ${\cal S}_i$ and ${\cal A}^\mu_i$
(see eqs. (\ref{scov}) and (\ref{acov})),
upper components have odd indices, lower components
have even indices.   

The strategy shown above for the octet baryon can also be applied to
decuplet baryons which have spin $3/2$.
Projection of the tri-spinor wave function
$\psi_{\alpha\beta\gamma}$ onto total spin 3/2 may be achieved
by expanding the piece associated with quark $a$ 
in terms of Rarita-Schwinger spinors $u_\mu$:
\be \label{3qstatedec}
 \psi^{10}_{\alpha\beta\gamma}=(\Gamma u_\mu)_\alpha(\phi C)_{\beta\gamma},
\ee
$\Gamma$ is chosen such that all Lorentz indices are contracted
with the momenta $P$ and $p$. Note that the Rarita-Schwinger
constraints demand $u_\mu \gamma^\mu = u_\mu P^\mu =0$ and that
in the diquark-quark model the diquark part of decuplet states
is made of axialvecor diquarks only, $\phi=C\gamma^\nu$.

The covariants $\Gamma$ which are left for decuplet states
after applying the above restrictions and the positive parity
constraint can be found in table \ref{covtable}, third row.
The covariants for octet and decuplet states are closely related. Due to 
the Rarita-Schwinger constraints the Lorentz index of $u_\mu$ must
be contracted either with the Lorentz index of the axialvector diquark
$(\gamma^\mu C)$ or with the transversal relative 
momentum
$p^\mu_T=p^\mu-\hat P^\mu (p\cdot\hat P)$.
For the first choice, we have the same four covariants as for
the octet state with scalar diquark correlations. For the second choice,
we obtain
the twelve covariants as for the octet state with axialvector diquark
correlations, however, multiplied by $\gamma_5$ to ensure positive parity. 

\newpage
The decuplet wave functions reads now:
\be \label{dec1}
 \psi^{10}_{\alpha\beta\gamma}&=&(\Upsilon^{\mu\nu}_{10} u^\nu)_\alpha 
(\gamma_\mu C)_{\beta\gamma} \\
 &=&\sum_{i=1}^4(Z_i^D \Gamma_i^S\delta^{\mu\nu}u^\nu)_\alpha 
(\gamma_\mu C)_{\beta\gamma}
 +\sum_{i=5}^{16}(Z_i^D \gamma_5\Gamma_i^{A\mu}\hat p^\nu_T u^\nu)_\alpha 
(\gamma_\mu C)_{\beta\gamma} \nonumber,
\ee
where the $Z_i^D$ denote the sixteen scalar functions for the decuplet state. 

Taking from eq. (\ref{dec1}) only the part with spinor index $\alpha$
and projecting it to positive energies leads to
\be
\sum_{s=-3/2}^{+3/2} \Upsilon^{\mu\nu}_{10}(p;P) 
u^\nu(P,s)\bar{u}^\lambda(P,s)
=\Upsilon^{\mu\nu}_{10}(p;P)
{\bf P}^{\nu\lambda}=
\Psi_{10}^{\mu \nu} (p;P)
= \Psi_{10}^{\mu \lambda} (p;P) {\bf P}^{\lambda \nu}. 
\ee
Therefore the projected wave function (which besides being
a 4$\times$4-matrix has tensor character) is determined by the
condition
\be
\Psi_{10}^{\mu \nu} (p;P) =
\Psi_{10}^{\mu \lambda} (p;P) {\bf P}^{\lambda \nu}, 
\label{RScond}
\ee
which requires $\Psi_{10}^{\mu \nu}$ to be an eigenfunction
of the Rarita-Schwinger projector.
Here, the explicit expression of the Euclidean Rarita-Schwinger
projector is given by

\be \label{RSP}
{\bf P}^{\mu\nu}:=\Lambda^+\left(-\delta^{\mu\nu}
+\frac{1}{3}\gamma^\mu\gamma^\nu-
\frac{2}{3} \frac{P^\mu P^\nu}{M^2} + 
\frac{i}{3} \frac{P^\mu\gamma^\nu-P^\nu\gamma^\mu}{M} \right) 
= :\Lambda^+ \Lambda^{\mu \nu}.
\ee  
\vspace{2mm}
The most general form which fulfills condition (\ref{RScond})
is 
\be
\Psi_{10}^{\mu \nu} (p;P) =D_{i}{\cal S}_i\Lambda^+\Lambda^{\mu\nu}+
iE_{i}\gamma_5 {\cal A}_i^\mu \Lambda^+ \hat p^\lambda_T \Lambda^{\lambda\nu}
\label{DELGEN}
\ee
which requires the covariants again to be eigenfunctions of $\Lambda^+$,
as a consequence we may use the same choice
for the ${\cal A}^\mu_i$ and ${\cal S}_i$ as in the octet case and are left
with eight independent scalar functions.

In the rest frame of the bound state, the decuplet wave functions are
then denoted by 

\begin{eqnarray}
\Psi^{ij}_{10} (p;P) &=&
   \pmatrix{(\delta^{ij}-\frac{1}{3}\sigma^i\sigma^j)\,D_1 & 0 \cr
    \frac{1}{p}({\vect{\sigma}}\vect{p})(\delta^{ij}-\frac{1}{3}\sigma^i\sigma^j)\,D_2 & 0\cr}
    + \nonumber \\
    & & \nonumber \\
   & & +  \pmatrix{(
- \hat p^i \, E_4 + i(\vect{\sigma} \times \hat{\vect{p}})^i \,E_6)
(\hat p^j-\frac{1}{3}(\vect{\sigma}\hat{\vect{p}})\sigma^j)& 0 \cr
\frac{1}{p}(-\hat p^i \, E_3 + i(\vect{\sigma} 
\times \hat{\vect{p}})^i \, E_5) 
(\vect{\sigma}\vect{p}) (\hat p^j-\frac{1}{3}
(\vect{\sigma}\hat{\vect{p}})\sigma^j)& 0 \cr}
\label{expdel} \\
 & & \nonumber \\
\Psi_{10}^{4j}(p;P) & = &
 \pmatrix{\frac{i}{p} (p^j-\frac{1}{3}(\vect{\sigma}\vect{p})\sigma^j)\, 
E_2 & 0 \cr
   i(\vect{\sigma} 
\hat{\vect{p}})(\hat p^j-
\frac{1}{3}(\vect{\sigma}\hat{\vect{p}})\sigma^j)\,E_1 & 0 \cr}
\nonumber 
\end{eqnarray}
and all other components of $\Psi_{10}^{\mu \nu}(p;P)$
vanish. The appearance of $\gamma_5$ in eq. (\ref{DELGEN})
has interchanged upper and lower components
as compared to the octet case.

Instead of working with the Bethe-Salpeter wave functions
one may alternatively use the Bethe-Salpeter
vertex functions obtained by amputating 
the external quark and diquark propagators from the wave function:
\be
\Phi_8 (p;P) & = & S^{-1}(p_a) D^{-1}(p_b) \Psi_8 (p;P), 
\label{Phidef} \\
\Phi^\mu_8 (p;P) & = & S^{-1}(p_a) (D^{-1})^{\mu\nu} 
(p_b) \Psi_8^\nu (p;P),
\label{Phimudef} \\
\Phi^{\mu\rho}_{10} (p;P) & = & 
S^{-1}(p_a) (D^{-1})^{\mu\nu} (p_b) \Psi^{\nu\rho}_{10} (p;P),
\label{DelPhimudef}
\ee
Substituting the wave functions by the vertex functions
in the Bethe-Salpeter equations (\ref{sc},\ref{deltaBSE})
leads to a reformulation of the bound state equations
which is sometimes
more convenient. For example in ref. \cite{Kus97,Hel97} the equations
containing only scalar diquarks have been solved in a form which
includes the vertex function explicitly. 

\subsection{Orbital Angular Momentum and Spin of the Bethe-Salpeter Wave Functions
in the Rest Frame of the Bound State} \label{LSeig}

Whereas the choice of the covariants in (\ref{scov},\ref{acov})
which build the octet and decuplet wave functions is well suited for numerical
computation and further covariant calculations, their physical
interpretation is not obvious. 

In general, covariant wave functions possess 
only the mass of the bound state $M$ and its total angular momentum $J$
as good quantum numbers.
In the rest frame of the bound state, however, the wave functions can be written
as a sum of tri-spinors each possessing definite orbital angular momentum
and spin, thus allowing a direct interpretation of the different components.
These tri-spinors are linear combinations of the covariants 
${\cal S}_i$ and ${\cal A}_i$ which have been constructed in the previous
subsection, multiplied by the respective Dirac matrices $(\gamma_5 C)_
{\beta\gamma}$ and $(\gamma^\mu C)_{\beta\gamma}$ denoting the diquark
content, respectively.

In the rest frame the Pauli-Lubanski operator for a tri-spinor is given by
\be
 W^i=\frac{1}{2}\epsilon_{ijk}{\cal L}^{jk},
\ee
whose square characterises the total angular momentum
\be
 W^iW^i\psi_{\alpha\beta\gamma}=J(J+1)\psi_{\alpha\beta\gamma}.
\ee
Here, $\psi_{\alpha\beta\gamma}$ is the tri-spinor wave function with positive 
parity and positive energy. 
The tensor ${\cal L}^{jk}$ is the sum of an orbital part, $L^{jk}$, and
a spin part, $S^{jk}$, which read
\begin{eqnarray}
 L^{jk}&=&\sum_{a=1}^3 (-i)\left(p_a^j\frac{\partial}{\partial p_a^k}-
         p_a^k\frac{\partial}{\partial p_a^j}\right), \\
 2(S^{jk})_{\alpha\alpha',\beta\beta',\gamma\gamma'}&=&
 (\sigma^{jk})_{\alpha\alpha'}\otimes{\delta}_{\beta\beta'}\otimes{\delta}_{\gamma\gamma'}+
 {\delta}_{\alpha\alpha'}\otimes (\sigma^{jk})_{\beta\beta'}
 \otimes{\delta}_{\gamma\gamma'}+ \nonumber \\
  & & {\delta}_{\alpha\alpha'}\otimes{\delta}_{\beta\beta'}\otimes(\sigma^{jk})_{\gamma\gamma'},  
\end{eqnarray}        
such that ${\cal L}^{jk}=L^{jk}+\frac{1}{2}S^{jk}$. 
Obviously, $L^{jk}$ is proportional to the unit matrix in Dirac space. 
The definition of $\sigma^{\mu\nu}:=-\frac{i}{2}[\gamma^\mu,\gamma^\nu]$ differs
by a minus sign from its Minkowski counterpart.
The tensors $L$ and $S$ are written
as a sum over the respective tensors for each of the three constituent quarks
which are labelled $a=1\dots 3$ and with respective Dirac indices $\alpha\alpha',
\beta\beta',\gamma\gamma'$.

With the definition
of the spin matrix $\Sigma^i=\frac{1}{2}\epsilon_{ijk}\sigma^{jk}$ the Pauli-Lubanski
operator reads
\begin{eqnarray}
(W^i)_{\alpha\alpha',\beta\beta',\gamma\gamma'} &=& L^i\,
  {\delta}_{\alpha\alpha'}\otimes{\delta}_{\beta\beta'}\otimes\delta_{\gamma\gamma'}
  +(S^i)_{\alpha\alpha',\beta\beta',\gamma\gamma'},  \\
L^i    &=&(-i)\epsilon_{ijk}p^j\frac{\partial}{\partial p^k}, \\
(S^i)_{\alpha\alpha',\beta\beta',\gamma\gamma'}   &=& \frac{1}{2} \left( 
   (\Sigma^{i})_{\alpha\alpha'}\otimes{\delta}_{\beta\beta'}\otimes
    {\delta}_{\gamma\gamma'}+
 {\delta}_{\alpha\alpha'}\otimes (\Sigma^{i})_{\beta\beta'}\otimes
 {\delta}_{\gamma\gamma'}+ \right. \nonumber \\
    & & \left.{\delta}_{\alpha\alpha'}\otimes{\delta}_{\beta\beta'}\otimes
 (\Sigma^{i})_{\gamma\gamma'}
 \right),
\end{eqnarray}
where we have already introduced the relative momentum $p$ between quark and diquark
via a canonical transformation:
\be
P=p^1+p^2+p^3,\quad p=\eta (p^1+p^2)-(1-\eta)p^3, \quad p'=\frac{1}{2}(p^1-p^2).
\ee
Assuming a pointlike diquark, the relative momentum between quark 1 and 2,
$p'$, vanishes and the only 
contribution to the orbital angular momentum stems from $p$.
${\bf W}^2$ now takes the form
\begin{eqnarray}
 {\bf W}^2&=& {\bf L}^2+2 \vect{L} \cdot \vect{S} + {\bf S}^2,   \\
{\bf L}^2 &=& \left(2p^i\frac{\partial}{\partial p^i}-\vect{p}^{2}\Delta_{p}+p^ip^j\frac{\partial}{\partial p^i}
          \frac{\partial}{\partial p^j}\right),  \\
 2 ({\bf L}\cdot{\bf S})_{\alpha\alpha',\beta\beta',\gamma\gamma'}&=& 
   -\epsilon_{ijk}p^j\frac{\partial}{\partial p^k} \left(
    (\Sigma^{i})_{\alpha\alpha'}\otimes{\delta}_{\beta\beta'}
     \otimes{\delta}_{\gamma\gamma'}+\right. \nonumber \\
  & &\left.{\delta}_{\alpha\alpha'}\otimes \left[(\Sigma^{i})_{\beta\beta'}
     \otimes{\delta}_{\gamma\gamma'}+
     {\delta}_{\beta\beta'}\otimes(\Sigma^{i})_{\gamma\gamma'}\right]\right),    \\
  ({\bf S}^2)_{\alpha\alpha',\beta\beta',\gamma\gamma'}  &= &
   \frac{1}{4}\left(9\,{\delta}_{\alpha\alpha'}\otimes{\delta}_{\beta\beta'}
  \otimes{\delta}_{\gamma\gamma'}
 +2\,{\delta}_{\alpha\alpha'}\otimes \left[(\Sigma^{i})_{\beta\beta'}\otimes{\delta}_{\gamma\gamma'}+
 {\delta}_{\beta\beta'}\otimes(\Sigma^{i})_{\gamma\gamma'}\right]\right. + \nonumber \\
    & & \left. +2\,{\delta}_{\alpha\alpha'}\otimes(\Sigma^{i})_{\beta\beta'}\otimes 
 (\Sigma^{i})_{\gamma\gamma'}\right).         
\end{eqnarray}

\begin{table}[t]
\begin{tabular}{|l|c|c|} \hline
$\psi^8_{\alpha\beta\gamma}$ in the rest frame & eigenvalue & eigenvalue \\ 
  & $l(l+1)$ of ${\bf L}^2$ & $s(s+1)$of ${\bf S}^2$ \\ \hline \hline
  & & \\
${\cal S}_1 u (\gamma_5 C)=\pmatrix{\chi \cr 0 \cr}(\gamma_5 C)$ & 0 & $\frac{3}{4}$ \\
${\cal S}_2 u (\gamma_5 C)=\pmatrix{0\cr \frac{1}{p}(\vect{\sigma}\vect{p})\chi \cr}(\gamma_5 C)$ & 2 & $\frac{3}{4}$ \\
 & & \\ 
${\cal A}^\mu_{1} u (\gamma^\mu C)=\hat P^0\pmatrix{\frac{1}{p}(\vect{\sigma}\vect{p})\chi\cr 0\cr}(\gamma^4 C)$ & 2 &$\frac{3}{4}$\\
${\cal A}^\mu_{2} u (\gamma^\mu C)=\hat P^0\pmatrix{0\cr\chi\cr}(\gamma^4 C)$ & 0 & $\frac{3}{4}$ \\
 & & \\
${\cal B}^\mu_{1} u (\gamma^\mu C)=\pmatrix{i\sigma^i\chi \cr 0 \cr}(\gamma^i C)$ & 0 & $\frac{3}{4}$ \\
${\cal B}^\mu_{2} u (\gamma^\mu C)=\pmatrix{0\cr\frac{i}{p}\sigma^i(\vect{\sigma}\vect{p})\chi\cr}(\gamma^i C)$ & 2 & $\frac{3}{4}$ \\ 
 & & \\
${\cal C}^\mu_{1} u (\gamma^\mu C)=\pmatrix{i\left(\hat p^i(\vect{\sigma}\hat\vect{p})-\frac{1}{3}\sigma^i\right)\chi\cr
         0\cr}(\gamma^i C)$ & 6 & $\frac{15}{4}$ \\ 
${\cal C}^\mu_{2} u (\gamma^\mu C)=\pmatrix{0\cr \frac{i}{p}\left(p^i-\frac{1}{3}\sigma^i(\vect{\sigma}\vect{p})\right)\chi\cr}
               (\gamma^i C)$ & 2 & $\frac{15}{4}$ \\ 
  & & \\             
\hline
\end{tabular}
\caption{\sf Classification of the components of the octet wave function
in terms of eigenstates of ${\bf L}^2$ and ${\bf S}^2$ in the rest frame of the
bound state. \label{octetLS}}
\end{table}

\begin{table}[t]
\begin{tabular}{|l|c|c|} \hline
$\psi^{10}_{\alpha\beta\gamma}$ in the rest frame & eigenvalue & eigenvalue \\ 
  & $l(l+1)$ of ${\bf L}^2$ & $s(s+1)$of ${\bf S}^2$ \\ \hline \hline
  & & \\
${\cal D}_1 u_\mu (\gamma^\mu C)=\pmatrix{\chi^i \cr 0 \cr}(\gamma^i C)$ & 0 & $\frac{15}{4}$ \\
${\cal D}^{\mu\nu}_2 u_\nu (\gamma^\mu C)=\pmatrix{0\cr \frac{1}{p}\left((\vect{\sigma}\vect{p})\chi^i 
-\frac{2}{3}\sigma^i(\vect{p}\vect{\chi})\right)\cr} (\gamma^i C)$ & 2 & $\frac{15}{4}$ \\ 
 & & \\
${\cal E}^{\mu\nu}_{1}u_\nu (\gamma^\mu C)=i\hat P^0\pmatrix{\frac{1}{p}(\vect{p}\vect{\chi})\cr 0\cr}(\gamma^4 C)$ & 2 &$\frac{3}{4}$\\
${\cal E}^{\mu\nu}_{2} u_\nu  (\gamma^\mu C)=i\hat P^0\pmatrix{0\cr (\vect{\sigma}\hat\vect{p})(\hat\vect{p}\vect{\chi}) \cr}
(\gamma^4 C)$ & 6 & $\frac{3}{4}$ \\
 & & \\
${\cal E}^{\mu\nu}_{3}u_\nu  (\gamma^\mu C)=\pmatrix{\sigma^i(\vect{\sigma}\hat\vect{p})(\hat\vect{p}\vect{\chi}) \cr
 0 \cr}(\gamma^i C)$ & 6 & $\frac{3}{4}$ \\
${\cal E}^{\mu\nu}_{4} u_\nu (\gamma^\mu C)=\pmatrix{0\cr\frac{1}{p}\sigma^i(\vect{p}\vect{\chi})\cr}
  (\gamma^i C)$ & 2 & $\frac{3}{4}$ \\ 
 & & \\
${\cal E}^{\mu\nu}_{5} u_\nu (\gamma^\mu C)=\pmatrix{\hat p^i(\hat\vect{p}\vect{\chi})-\frac{1}{3}[\chi^i
   +\sigma^i (\vect{\sigma}\hat\vect{p})(\hat\vect{p}\vect{\chi})]   \cr
         0\cr}(\gamma^i C)$ & 6 & $\frac{15}{4}$ \\ 
${\cal E}^{\mu\nu}_{6} u_\nu (\gamma^\mu C)=\pmatrix{0\cr \frac{1}{p}\left(p^i(\vect{\sigma}\hat\vect{p})(\hat\vect{p}\vect{\chi})
  -\frac{1}{5}[\sigma^i(\vect{p}\vect{\chi})+
   (\vect{\sigma}\vect{p})\chi^i]\right) \cr }
               (\gamma^i C)$ & 12 & $\frac{15}{4}$ \\ 
  & & \\             
\hline
\end{tabular}
\caption{\sf Classification of the components of the decuplet wave function
in terms of eigenstates of ${\bf L}^2$ and ${\bf S}^2$ in the rest frame of the
bound state. \label{decLS}}
\end{table}

When applying ${\bf W}^2$ to the wave functions we first note that a scalar function does not
contribute to the angular momentum, e.g.
\be
{\bf L}^2\,S_1(p^2, iMp_4)= 2\vect{L}\cdot\vect{S} \,S_1(p^2, iMp_4)=0.
\ee  

In table \ref{octetLS} we express all terms of the octet wave function in 
terms of eigenstates of 
${\bf L}^2$ and ${\bf S}^2$. To this end we define the
following linear combinations of matrices:
\begin{eqnarray}
{\cal B}_1^\mu&=&{\cal A}_5^\mu + {\cal A}_3^\mu, \nonumber \\
{\cal B}_2^\mu&=&{\cal A}_6^\mu + {\cal A}_4^\mu,\nonumber \\
{\cal C}_1^\mu&=&- \frac{1}{3}{\cal A}_5^\mu+\frac{2}{3}{\cal A}_3^\mu, \nonumber \\
{\cal C}_2^\mu&=&- \frac{1}{3}{\cal A}_6^\mu+\frac{2}{3}{\cal A}_4^\mu .
\end{eqnarray}
The eigenvalue of ${\bf W}^2$ is $\frac{3}{4}$ for all terms, of course. 
In the table, $\chi=\{\chi^+,\chi^-\}$ denotes
an arbitrary Pauli two-component spinor which is the positive energy basis
for quark $a$ with Dirac index $\alpha$. 
To derive this, the following relations between Dirac matrices
have proven to be useful,
\begin{eqnarray}
\Sigma^j(\gamma_5 C)+(\gamma_5 C)(\Sigma^j)^T&=&0 \\
 & & \nonumber \\
\Sigma^j\pmatrix{(\gamma^4 C) \cr (\gamma^i C) \cr}+\pmatrix{(\gamma^4 C) \cr (\gamma^i C) \cr}
   (\Sigma^j)^T &=& \pmatrix{ 0 \cr 2i\epsilon_{mji} (\gamma^m C)\cr}, \\
   & & \nonumber \\
\Sigma^j (\gamma_5 C) (\Sigma^j)^T &=& -3 (\gamma_5 C), \\
& & \nonumber \\
\Sigma^j \pmatrix{(\gamma^4 C) \cr (\gamma^i C) \cr} (\Sigma^j)^T &=& \pmatrix{-3(\gamma^4 C) 
    \cr (\gamma^i C) \cr}.
\end{eqnarray}

Three covariants can be regarded as ``$s$-wave'' components, ${\cal S}_1$, ${\cal A}_2$ and 
${\cal B}_1$, and we expect the corresponding scalar functions
to dominate the wave function decomposition.
The other ``$p$-, $d$-wave'' components represent all remaining possibilities of combining
orbital angular momentum between quark and diquark and the joint spin of axialvector diquark
and quark to total spin 1/2. In this sense the description is closed. 

The individual terms of the decuplet wave function can be classified accordingly
with the help of the linear combinations:
\begin{eqnarray}
{\cal D}_1&=&{\cal S}_1, \nonumber \\
{\cal D}^{\mu\nu}_2&=&{\cal S}_2 \delta^{\mu\nu}+
               \frac{2i}{3}\gamma_5({\cal A}^\mu_5+{\cal A}^\mu_3)\hat p^\nu_T,\nonumber  \\
{\cal E}^{\mu\nu}_1&=&i\gamma_5 {\cal A}_2^\mu \hat p^\nu_T,\nonumber \\
{\cal E}^{\mu\nu}_2&=&i\gamma_5 {\cal A}_1^\mu \hat p^\nu_T,\nonumber \\
{\cal E}^{\mu\nu}_3&=&-i\gamma_5({\cal A}_6^\mu+{\cal A}_4^\mu) \hat p^\nu_T,\nonumber \\
{\cal E}^{\mu\nu}_4&=&-i\gamma_5({\cal A}_5^\mu+{\cal A}_3^\mu) \hat p^\nu_T,\nonumber \\
{\cal E}^{\mu\nu}_5&=&\left(-\frac{2i}{3}\gamma_5{\cal A}_4^\mu+\frac{i}{3}\gamma_5
         {\cal A}_6^\mu\right)\hat p^\nu_T -\frac{1}{3}{\cal S}_1\delta^{\mu\nu},\nonumber \\
{\cal E}^{\mu\nu}_6&=&\left(-\frac{4i}{5}\gamma_5{\cal A}_3^\mu+\frac{i}{5}\gamma_5
         {\cal A}_5^\mu\right)\hat p^\nu_T -\frac{1}{5}{\cal S}_2\delta^{\mu\nu}.
\end{eqnarray}
The result can be found in table \ref{decLS}.         
Due to the spin projection by use of 
the Rarita-Schwinger projector the eigenvalue of ${\bf W}^2$ is $\frac{15}{4}$. 
$\chi^i=\{\chi^{+i},\chi^{-i}\}$ ($i$=1,2,3)  denotes
a two-component vector-spinor which survives the spin-3/2 projection in the rest frame. The
Rarita-Schwinger constraints reduce to $\vect{\sigma}\vect{\chi}=0$.

For the decuplet only one $s$-wave component exists, ${\cal D}_1$. 
Again these covariants exhaust all possible couplings of
spin and orbital angular momentum, note that even an orbital angular
momentum $l$=3 contributes to a spin-3/2 state, ${\cal E}_6$.
A contribution of an $l$=1 state with scalar diquark, such as
$\pmatrix{0\cr (\vect{p}\vect{\chi}) \cr}(\gamma_5 C)$,  is forbidden by the Pauli
principle for pure decuplet states, but may admix in the case of broken flavour
symmetry.

\section{Ladder approximation to the Bethe-Salpeter equation}
\label{Rootchap}
In our preceding publication \cite{Hel97} we used a momentum 
routing in the Bethe-Salpeter equation, where the interaction
(propagator of the exchanged quark)
is manifestly independent of the total momentum $P$ of the baryon,
i.e. $q=-p-p^\prime$.
A $P$-independent kernel is desirable for the following reasons:
As will be discussed in the next section, it  reduces the 
numerical work when solving the integral equation quite  
drastically. Furthermore, the canonical normalisation
condition for the Bethe-Salpeter wave function \cite{Itz85}
becomes much more involved when the interaction is $P$-dependent.
Therefore, when solving a Bethe-Salpeter equation
in ladder approximation one usually 
tries to find  a momentum routing 
having  this very convenient feature. 
In our equations, however, where quark and diquark interact through
quark exchange, which changes the identity of the particles
after each interaction (a diquark becomes a quark and a quark 
becomes a diquark after the quark exchange), demanding 
a $P$-independent interaction defines a ``modified 
ladder approximation'' with results deviating from other
momentum routings. Note, that this problem does not occur
in the Bethe-Salpeter approach if  the interaction,
e.g. meson exchange in the two nucleon system or gluon 
exchange between quark and antiquark, 
does not change the identity of the constituents.
Almost all Bethe-Salpeter equations treated in the literature are 
of this kind. For a comprehensive review of the existing
Bethe-Salpeter literature see \cite{Nak88}.
In order to clarify the ambiguity arising in the diquark-quark
Bethe-Salpeter equation, which came up during
our investigations, we discuss here two possible 
momentum routings and in section 6 and appendix \ref{Procadis}
the corresponding variations of the results.

For sake of clarity of this 
discussion we temporarily restrict ourselves
to the part of the octet equation which involves only
scalar diquarks, i.e. eq. (\ref{sc}) with $g_a=0$.  
Furthermore, it is more  transparent to work with the Bethe-Salpeter
equation which involves the vertex function (see eq. (\ref{Phidef})):
\be
\Phi_8(p;P) &= & - |g_{s}|^{2} 
\int \frac{d^4p^{\prime}} {(2\pi)^4}
\gamma _5 \tilde S(-q) \gamma _5 
S(p_{a}^\prime) D(p_{b}^\prime)  
\Phi_8(p^{\prime};P).
\label{scex}   
\ee 
Repeating the procedure discussed in section \ref{Ansatzchap}
for the vertex instead of the wave function leads to an ansatz \cite{Kus97,Hel97}
\be
\Phi_8(p;P)=
\left( \begin{array}{cc}
{\bf 1} \, \hat{S}_1(p,z) & 0 \\
\frac{1}{p}(\vect{\sigma}\vect{p})\, \hat{S}_2(p,z) & 0 \\ 
\end{array}\right).
\ee
Note that
in the lower component of this spinor-like object 
the spin is orientated along the
spatial part of the relative momentum.
In the Bethe-Salpeter equation (\ref{scex})
the momentum of the constituent quark
is given by $p_a^\prime =p^\prime+\eta P$, and the momentum
of the scalar diquark by  $p_b^{\prime}=-p^{\prime}+(1-\eta) P$.
$P$ denotes the 
total momentum of the nucleon 
and $p$ the relative momentum between quark and diquark.
The Mandelstam parameter $\eta$ 
describes how $P$ is partitioned 
to quark and diquark. The fact that  the eigenvalues of the Bethe-Salpeter equation
are independent of $\eta$
is a direct consequence of Lorentz covariance.

\begin{figure}[t]
\centerline{\epsfxsize 11.0cm
\epsfbox{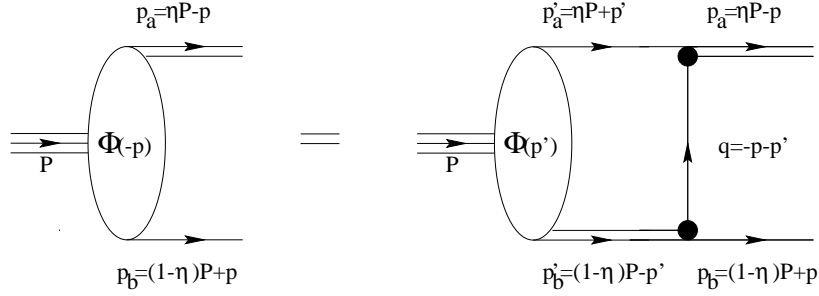}
}
\caption{\sf Momentum routing defining the modified ladder approximation.
\label{modrout}}
\end{figure}

If the momentum of the exchanged quark is chosen 
as $q=-p-p^{\prime}$, the interaction kernel is by construction 
independent
of the total momentum.
Since the quark exchange transforms a quark to a diquark and vice versa,
after the interaction, the relative momentum between these
particles becomes $(1-\eta)p_a-\eta p_b=-p$, see fig. \ref{modrout}
for this momentum routing. We now  demand
$\Phi_8(-p;P)=\Phi_8(\vect{p},-z,P)$, 
i.e. the orientation
of the nucleon spin should not depend  on the orientation 
of $p$ as required by the Dirac decomposition before projection 
onto positive energies. 
Then the vertex function appearing on the left hand side
in eq. (\ref{scex}) reads 
\be
\Phi_8(p;P)=
\left( \begin{array}{cc}
{\bf 1} \, \hat{S}_1(p,-z) & 0 \\
\frac{1}{p}(\vect{\sigma}\vect{p})\, \hat{S}_2(p,-z) & 0\\ 
\end{array}\right).
\label{moddef}
\ee
The definition of the momentum $q$ together with eq. (\ref{moddef})
is what we call ``modified ladder approximation'' which leads 
to satisfying results for various nucleon form factors in the weak binding
regime, as can 
be seen in ref. \cite{Hel97}. Using such a  
prescription we have been able to reproduce the results of
ref. \cite{Kus97}.

\begin{figure}[b]
\centerline{\epsfxsize 11.0cm
\epsfbox{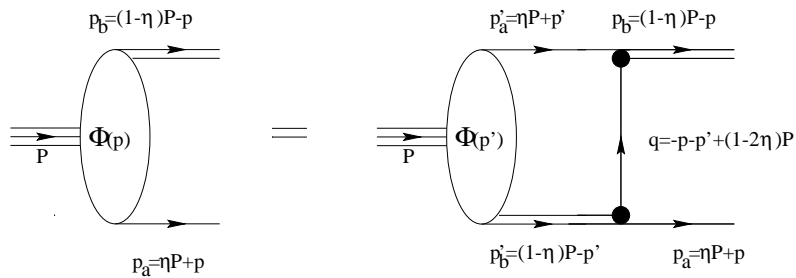}
}
\caption{\sf Momentum routing defining the direct ladder approximation.
\label{dirrout}}
\end{figure}

Another possible momentum routing allows $P$ to flow through
the quark exchange,
$q=-p-p^{\prime}+ (1- 2 \eta) P$, see fig. \ref{dirrout}. When taking not into account,
that quark and diquark change their role after exchanging a quark,
that means attributing the index ``a'' to the quark and index ``b'' to the diquark,
the relative momentum after the interaction is  
given by $p$.
Correspondingly, the vertex function on the left hand side in
eq. (\ref{scex}) is denoted by
\be
\Phi_8(p;P)=
\left( \begin{array}{cc}
{\bf 1} \, \hat{S}_1(p,z) & 0 \\
\frac{1}{p}(\vect{\sigma}\vect{p}')\, \hat{S}_2(p,z) & 0 \\ 
\end{array}\right).\\
\ee  
We name this ``direct ladder approach''.
Although one might  expect that both choices of the
momentum routing lead to the same physical results, we found
that this is actually not the case. 
In the next sections and appendix \ref{Procadis} we report on 
calculations using both momentum routings, see also \cite{Sme98}
for further results obtained in the direct ladder approach.
We found that 
physical results slightly differ for these two choices.
We will show, however, that both possibilities are manifestly Lorentz 
covariant, i.e. the eigenvalues do not depend
on the Mandelstam parameter $\eta$.
Furthermore, for weak binding the eigenvalues almost coincide.
Given that the ladder approximation is only reliable 
for weak binding we conclude that both methods are of similar 
validity.

Although during this discussion we restricted ourselves for clarity
to scalar diquarks and to the Bethe-Salpeter equation
involving the vertex function, it can be extended in a straightforward
way to the 
complete equation including the axialvector diquark channel,
the decuplet equation and also to the equations (\ref{sc},\ref{deltaBSE})
involving the wave functions
in an obvious way. The reported numerical results are always given
for the full problem. 
  
\section{Numerical method}
\label{Numchap}
For the numerical solutions of the Bethe-Salpeter equations
we developed an iterative hybrid algorithm, which allows
a very efficient and fast computation. A description of our 
numerical method can be found in a forthcoming publication
which presents this algorithm in all details \cite{Oet98}. 
Thus we will focus here on the main steps only.

We solve the Bethe-Salpeter equation as a system of equations for the
wave function $\Psi$ and the  
vertex function $\Phi$, see eq. (\ref{sc}, \ref{deltaBSE}). Both wave
and vertex function can be expanded in the rest frame according to \mbox{eq.
(\ref{expnuc}, \ref{expdel})}.
Although there are 10 equations for the eight octet functions, stemming
from 2 equations in the scalar diquark channel and 2$\times$4 equations
in the axialvector diquark channel, we confirmed that two of them are redundant.
The decuplet system yields 2$\times$12 equations for eight scalar functions due
to the tensor character of the wave function which reduce again to eight independent
equations. This especially underlines the necessity of including the subdominant
amplitudes describing orbital angular momentum to keep the system closed.

We expand the scalar functions (amplitudes) $S,A$ and $D,E$
defined in (\ref{NUCGEN},\ref{DELGEN}) into Chebyshev polynomials 
of the second kind,

\begin{eqnarray}
Y_i(P;p) &=& \sum_{n=0}^\infty i^n Y_i^n(P^2;p^2)U_n(\hat P \cdot \hat p) \\
\hat Y_i(P;p) &=& \sum_{n=0}^\infty i^n
      \hat Y_i^n(P^2;p^2)U_n(\hat P \cdot \hat p), 
\end{eqnarray}
where amplitudes with a hat, i.e. $\hat Y_i$, belong to the vertex function and the ones without
a hat to the wave function. Here we use the generic label $Y_i$ according to
 $Y_i=\{S_1,S_2, \dots, A_6\}_{octet}$, $\{D_1,D_2, \dots, E_6\}_{decuplet}$.

Throughout the calculation we work with usual hyperspherical coordinates,
\be
\pmatrix{p_1'\cr p_2'\cr p_3'\cr p_4' \cr}=p'
    \pmatrix{\sin\psi'\sin\theta'\sin\phi'\cr\sin\psi'\sin\theta'\cos\phi'\cr
             \sin\psi'\cos\theta'\cr \cos\psi'},
\quad z'=\cos\psi'=\hat P\cdot \hat p'. 
\ee
We are free to choose the spatial part of the relative momentum $p$ appearing
on the l.h.s. of the Bethe-Salpeter equation (\ref{sc}). Without loss of generality, we
select $p^\mu=(0,0,p\sqrt{1-z^2},pz)$.

We expand quark and diquark propagators into Chebyshev polynomials 
as well and project the Bethe-Salpeter equation onto the Chebyshev moments
of the amplitudes, $\hat Y_i^m, Y_i^m$. Note that in the chosen Lorentz frame 
with the relative momentum ${\bf p}$ 
parallel to the third axis this is especially
easy because in this case the amplitudes as given in (\ref{expnuc}) do not mix.
The integration necessary to generate the kernel will be performed in
 hyperspherical coordinates.
The integrations over $\phi'$ and $\theta'$
are done analytically, and the  remaining two over $z'$ and $z$ (due to the
projection) numerically.

The final equation suitable for iteration or diagonalisation reads:

\begin{eqnarray}
Y_i^m(p_{l_1}) &=& -g^2 \sum_{j=1}^{8} \sum_{n=0}^{n_{max}}
                P_{ij}^{mn}(p_{l_1}) \hat Y^n_j(p_{l_1}) \\
\hat Y^n_j(p_{l_1}) &=& \sum_{k=1}^{8} \sum_{m=0}^{m_{max}}
                        \int_0^\infty dp'\,p'^3 H_{jk}^{nm}(p_{l_1},p'_{l_2})
 Y_k^m(p'_{l_2}).
\end{eqnarray}

Here, $P_{ij}^{mn}$ is the propagator matrix with amplitude indices $i,j$
and Chebyshev moment indices $m,n$ and $H^{nm}_{jk}(p,p')$ is the
matrix of the quark exchange kernel given on a momentum grid $(p_{l_1},p'_{l_2})$
with respective amplitude indices $j,k$ and
Chebyshev moment indices $n,m$. The sum over amplitude indices
runs from 1\dots 8 for both octet and decuplet.
$m_{max}$ and $n_{max}$ denote the highest Chebyshev polynomial considered
in the expansion of the vertex and the wave function amplitudes, respectively. 
The kernel includes also the flavour factors and the
ratio $g_a/g_s$ (for spin-1/2 baryons). It is only in the 
modified ladder approximation that
the integral kernel does not include an explicit dependence 
on the bound state mass $M$ for all values of $\eta$, which makes a
fast determination of $M$ for given couplings feasible.
We refer to $g$ as the eigenvalue of the Bethe-Salpeter equation which is 
$g_s$ for spin-1/2
and $g_a$ for spin-3/2 baryons.

\begin{table}[t] \centerline{
\begin{tabular}{|lr|r|r|r|r|} \hline
 \multicolumn{6}{|c|}{Confining Propagators, $d$=10} \\ \hline
 & $n_{max}$ & 0 & 1 & 2 & 3 \\ 
$m_{max}$ & & & & & \\ \hline
0 & & 9.4362 & 9.1533 & 9.1417 & 9.1420  \\
1 & & 9.7480 & 9.2277 & 9.2260 & 9.2265 \\
2 & & 9.7568 & 9.2276 & 9.1992 & 9.1988 \\
3 & & 9.7568 & 9.2277 & 9.1994 & 9.1990  \\ \hline \hline
\multicolumn{6}{|c|}{Confining Propagators, $d$=1} \\ \hline
 & $n_{max}$ & 0 & 1 & 2 & 3 \\ 
$m_{max}$ & & & & & \\ \hline
 0 & & 11.3797 & 10.9507 & 10.9544 & 10.9547 \\
 1 & & 11.8555 & 11.1401 & 11.1704 & 11.1711 \\
 2 & & 11.8656 & 11.1503 & 11.1443 & 11.1445 \\
 3 & & 11.8656 & 11.1504 & 11.1444 & 11.1446 \\ \hline 
\end{tabular}} 
\caption{\sf Octet Bethe-Salpeter Equation - eigenvalues $g_s$ for various 
$m_{max}$ and $n_{max}$, the
maximum Chebyshev momenta of vertex and wave function. Parameters:
$m_q$=$m_{0^+}$=$m_{1^+}$, $M$=1.5$m_q$, $\Lambda$=2$m_q$,
$g_a/g_s$=0.5, $\eta$=0.5, momentum grid size $n_p$=20. 
\label{conv}}
\end{table}

In the following we will investigate the convergence properties of the 
expansion in Chebyshev polynomials. Furthermore, the independence
of the eigenvalue from the Mandelstam parameter $\eta$ will be discussed.
The chosen example is the octet equation (\ref{sc}) with the axialvector 
diquark propagator diagonal in Lorentz indices
and a diquark size factor of the monopole type. 
  
For all calculations reported, a momentum mesh size of 20$\times$20 
and the inclusion of only zeroth, first and 
second Chebyshev moment for both vertex and wave function amplitudes 
is sufficient for determining the eigenvalue
up to $10^{-4}$ precision (see table \ref{conv}). 
This is also reflected in the magnitudes of the vertex amplitudes $\hat Y_i$:
Going up one Chebyshev moment suppresses the amplitudes by almost one order of 
magnitude. The wave function amplitudes $Y_i$ converge somewhat slower, see
also sect. \ref{wf}.
For a large pole screening factor, $d>5$, and a weak binding situation the even Chebyshev 
momenta are more pronounced than the odd ones.

Increasing the coupling constant $g_a$ in the nucleon equation 
always lowers the eigenvalue, hence the axialvector diquark enhances
binding. 

\begin{figure}[t]
\centerline{\epsfxsize 7.5cm
\epsfbox{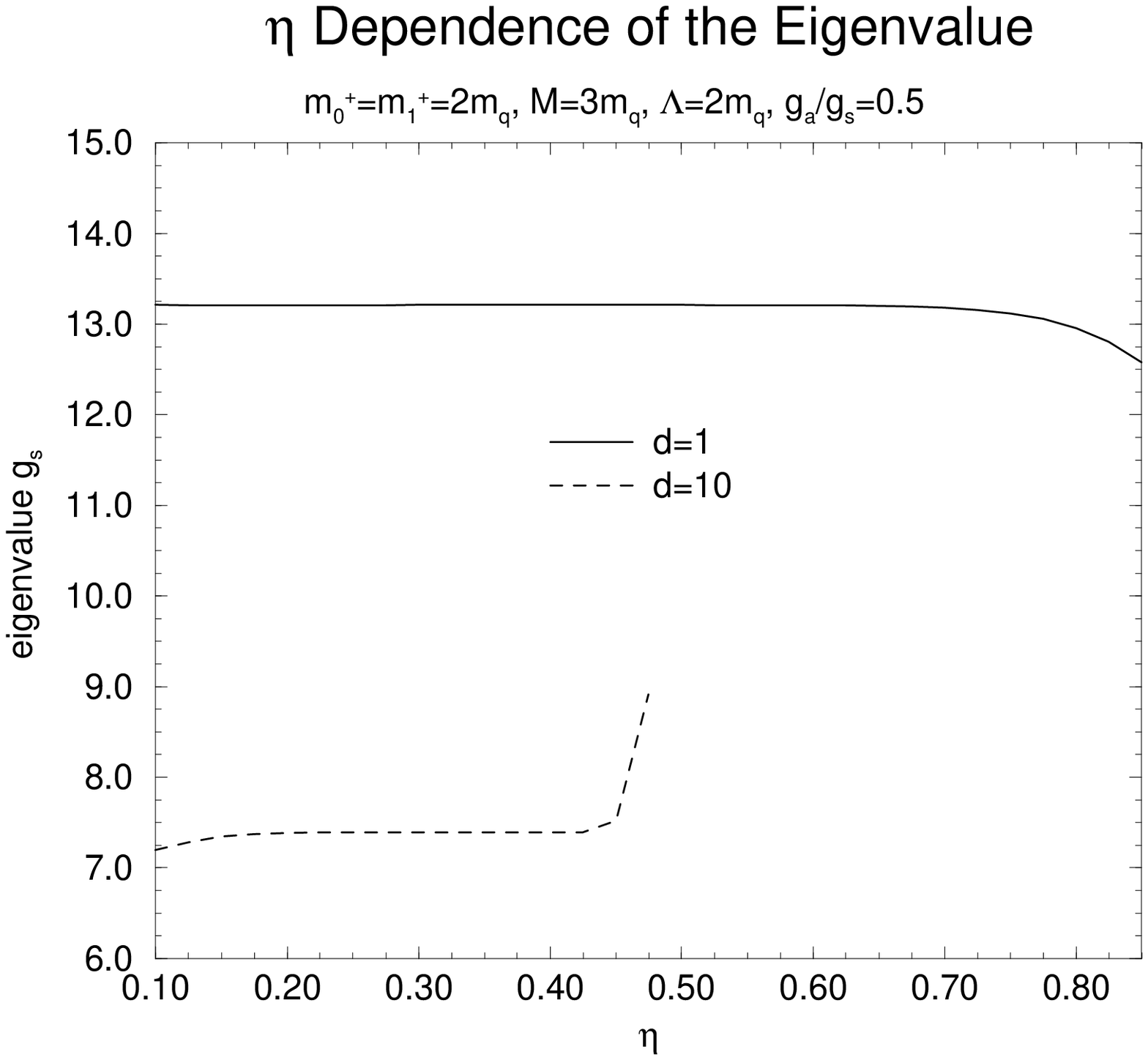}
\epsfxsize 7.5cm
\epsfbox{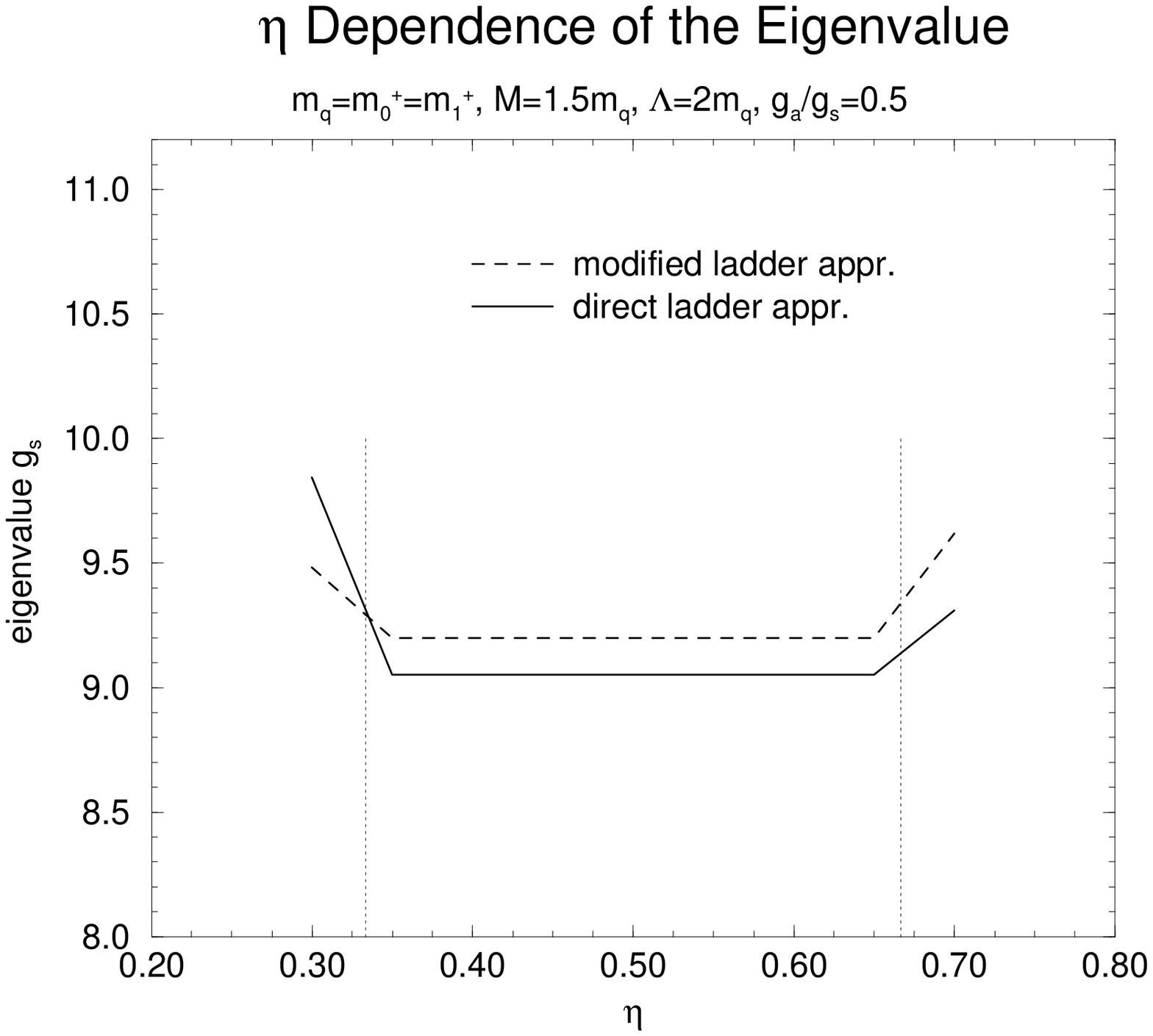}
}
\caption{\sf Eigenvalue vs. Mandelstam parameter $\eta$, left: 
two pole screening factors compared, right:
modified and direct ladder approach compared. Dotted
lines denote the location of the propagator poles.\label{eta}}
\end{figure}

For tree level propagators the choice of the Mandelstam parameter
$\eta$ is limited to the values $\eta \in [1-\mbox{min}\{m_{0^+},m_{1^+}\}/M,
m_q/M]$ to avoid singularities in the propagators. This restriction should not
apply for confining propagators. To demonstrate this, we choose
the extreme case $m_{0^+}$=$m_{1^+}$=2$m_q$ and the tree level threshold
$M$=3$m_q$\footnote{For similar choices
of quark and diquark masses we calculate the octet and decuplet
masses, see sect. \ref{Reschap}. The masses of the decuplet baryons
are close to the tree level threshold.}.
Then, tree level propagators limit the choice to $\eta$=1/3, as
opposed to confining propagators. For a small pole screening factor $d$=1 we 
could vary $\eta$ in a wide range without affecting
the eigenvalue, for $d$=10 due to numerical instabilities
the invariance region depends
on the constituent masses, displayed in figure \ref{eta} in the left panel.
It can be seen that
the poles in the tree level propagators are effectively screened although
the invariance region for $d$=10 is restricted to 0.2\dots0.42.

Whereas the latter results were obtained in the modified ladder approximation,
the check of $\eta$-independence for the direct ladder approach
requires the total momentum $P$ to appear in the kernel: $q=-p-p'+
(1-2\eta)P$, which slows down the numerics considerably. In the case of
tree level propagators, the right panel of figure \ref{eta} shows the invariance
of the eigenvalue for both momentum routings.

\begin{figure}[h]
\centerline{\epsfxsize 7.5cm
\epsfbox{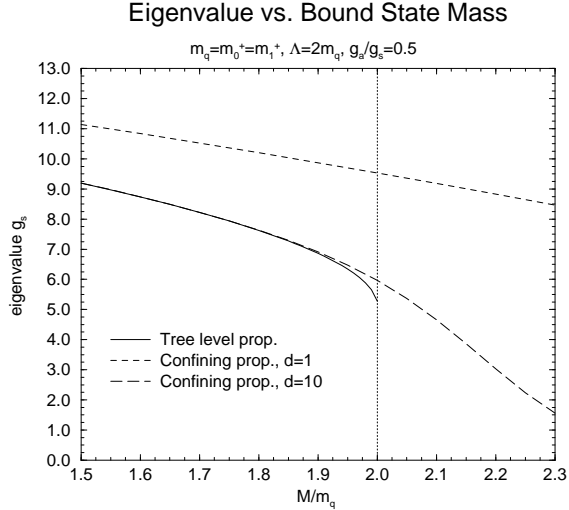}}
\caption{\sf Eigenvalues vs. bound state mass $M$. Absence of threshold 
effects for confining propagators.\label{thresh}}
\end{figure}

The crucial advantage of using confining propagators can be seen
even more clearly in figure \ref{thresh}.
Here the function $g_s(M)$ decreases rapidly for tree level propagators 
near threshold while the corresponding function for confining propagators
runs smoothly over the ``pseudo''-threshold. We furthermore observe that 
the even Chebyshev momenta of the
Bethe-Salpeter wave functions for tree 
level propagators close to threshold  become squeezed 
in the low momentum domain while the corresponding odd momenta are 
suppressed. No such effect is present for the Bethe-Salpeter wave functions
obtained with confining propagators.

All the numerical features described also apply to the decuplet
equation in the weak binding regime ($M>\frac{3}{4}(m_q+m_{0^+,1^+})$). 
For lower bound state masses the convergence becomes worse and
for $M<\frac{1}{2}(m_q+m_{0^+,1^+})$ our method failed in finding a real eigenvalue
for the coupling $g_a$. In this mass region an effect well known from
the Cutkosky model is visible: two states described by their functions
$g_a(M)$ collide and form a complex conjugate pair of eigenvalues \cite{Kau69,Ahl98}.
This shortcoming of the ladder approximation provides us with an upper limit
of the coupling constant $g_a$ for which the approximation is valid.
However, the $g_a$ needed to describe the baryon decuplet is well below
this critical value.

\section{Results: Masses and Selected Wave Functions}
\label{Reschap}

\begin{table}[t] \centerline{
\begin{tabular}{|l|l|l|l|l|} \hline
 & exp. &  \multicolumn{2}{|c|}{mod. ladder}& dir. ladder \\
 &       & I & II & III  \\ \hline
$d$       &      &10       &1     &1   \\
$\Lambda\,\,$ (GeV) &      &1         &1     &1  \\
$m_u\,$ (GeV)    &      &0.5    &0.5   &0.5  \\
$m_s$ (GeV) &      &0.65  &0.63  &0.63   \\ 
$\xi$     &      &1         &0.73  &0.73  \\ \hline
$g_a$     &      &10.35   &10.92  & 10.05\\
$g_s$     &      &9.43 &8.06 & 7.34 \\ \hline
$M_\Lambda\,$ (GeV) &1.116 &1.123 &1.130 &1.133 \\
$M_\Sigma\,$ (GeV) &1.193 &1.134 &1.137 &1.140 \\
$M_\Xi\,$ (GeV)     &1.315 &1.307 &1.319 &1.319 \\ \hline
$M_{\Sigma^*}$ (GeV)&1.384 &1.373&1.372 &1.380 \\ 
$M_{\Xi^*}$ (GeV)   &1.530 &1.545 &1.548 &1.516 \\
$M_\Omega\,$ (GeV)  &1.672 &1.692 &1.697 &1.665 \\ \hline \hline
 $\chi^2$ &      &0.0028  &0.0028 &0.0021 \\ \hline
\end{tabular}}
\caption{\sf Octet and decuplet masses obtained with the maximum 
order in Chebyshev
polynomials $m_{max}$=$n_{max}$=3 and momentum grid size $n_p$=20. 
\label{baryonmasses}}
\end{table}
\subsection{Octet and Decuplet Masses}

In our approach the strange quark constituent mass
$m_s$ is the only source of flavour symmetry breaking. Isospin
is assumed to be conserved. The equations describing octet 
and decuplet baryons have been derived under the
premises of flavour and spin conservation, i.e. only wave function components
with same spin and flavour content couple. The flavour structure of
the eight equations describing $N, \Lambda, \Sigma, \Xi, 
\Delta, \Sigma^*, \Xi^*$ and $\Omega$
can be found in appendix \ref{flavoureq}. 

In order to limit the number of parameters we assume  
the scalar and axialvector diquark masses to be equal.
Furthermore, we choose them to be $m^{fg}_{0^+,1^+}=\xi(m_f+m_g)$
where $fg \in \{uu,us,ss\}$ is the flavour content of the diquark.
We denote $\xi$ as diquark mass parameter. Assuming $\xi \in [0,1]$
is obviously natural.

With these choices the model
has the following parameters: two constituent quark masses $m_{u}=m_{d},
m_s$, the pole screening factor $d$, the diquark size factor $\Lambda$,
the diquark mass parameter $\xi$ and the couplings $g_a,g_s$.

We do not try to make a thorough fit onto the baryon masses in our 
parameter space. 
Inspired
by our results for the form factors of proton and neutron \cite{Hel97},
we assume first $d$=10. This results in only slight modifications  
of the propagators compared to the tree level ones for spacelike momenta.
Furthermore, we choose $\Lambda$=1 GeV $\approx 2m_u$ 
and $\xi$=1 to stay close to
calculations of octet and decuplet masses in NJL-diquark-models as 
done in \cite{Han95} and \cite{Buc95}. These authors, however,
used just a static approximation to the Bethe-Salpeter equation. 
Note that this part of our calculation has been done 
only in modified ladder approximation since we were forced to 
choose $\eta<$0.4 (cf. figure \ref{eta}), and solely for $\eta$=0.5
the integral kernel does not depend on the baryon mass in the direct approach,
which allows a fast determination of $M$.
We then fix the couplings $g_a$ and $g_s$ by
the nucleon and delta mass. The ratio $g_a/g_s$ is quite independent
from the ratio $M_\Delta/m_q$ and varies weakly with $d$.
Finally we vary just the two 
constituent quark masses to obtain the other six hyperon masses reasonably
close to their experimental values.

Secondly,
we explore the case $d$=1 and now additionally allow the diquark 
mass factor $\xi$ to vary. In \cite{Hel97} we showed that this 
strong screening of the propagator
poles leads to overestimated nucleon e.m. form factors for high $Q^2$. 
In this case we also compare
results for the two momentum routings, modified and direct ladder approach.
In both calculations $\eta$ is set to 0.5.

The results are given in table \ref{baryonmasses}. 
The quality of the results may be read from the quantity $\chi^2=
\sum_{i=1}^6\frac{(M^i_{theor}-M^i_{exp})^2}{(M_{exp}^i)^2}$.
Note that the mean deviation of the calculated masses from
the experimental ones is of the order of half a per cent or less.
Column I shows the results for $d$=10.
A remarkable feature is the large constituent quark mass $m_u$=500 MeV, whereas 
the constituent mass difference between strange and up quark is 150 MeV,
a commonly used value.
In the following two subsections we will show for this set of parameters
the wave function amplitudes for the nucleon and the delta
and discuss the contribution of the various components
with respect to their orbital angular momentum and exemplify flavour symmetry
breaking effects on the wave and vertex function of
the $\Lambda$-hyperon. 

The next columns, columns II and III, show the results for $d$=1
with the other parameters chosen to give baryon masses to the
same level of accuracy as before. 
In this case quark and
diquark propagators are strongly modified for spacelike momenta.
Quark and diquark ``masses'' begin to loose their meaning 
which we usually attach to them.
This is reflected in our solutions in a rather small diquark mass parameter
which renders constituent quarks and diquarks roughly equal in their masses.
Direct and modified ladder approach give approximately the same results
in the latter case. 

In all cases the Gell-Mann-Okubo mass formula for the octet is fulfilled
with an inaccuracy of less than 0.5\%. It should be emphasized that
the mass splitting between octet and decuplet is exclusively provided
by the quark exchange with the coupling strengths $g_s$ and $g_a$ and is not
due to a heavier axialvector diquark as assumed in other diquark-quark models,
e.g. in \cite{Vog91}.

\subsection{Vertex Functions for the $\Lambda$-hyperon}

\begin{figure}[t]
\centerline{{
\epsfxsize 5.9cm
\epsfbox{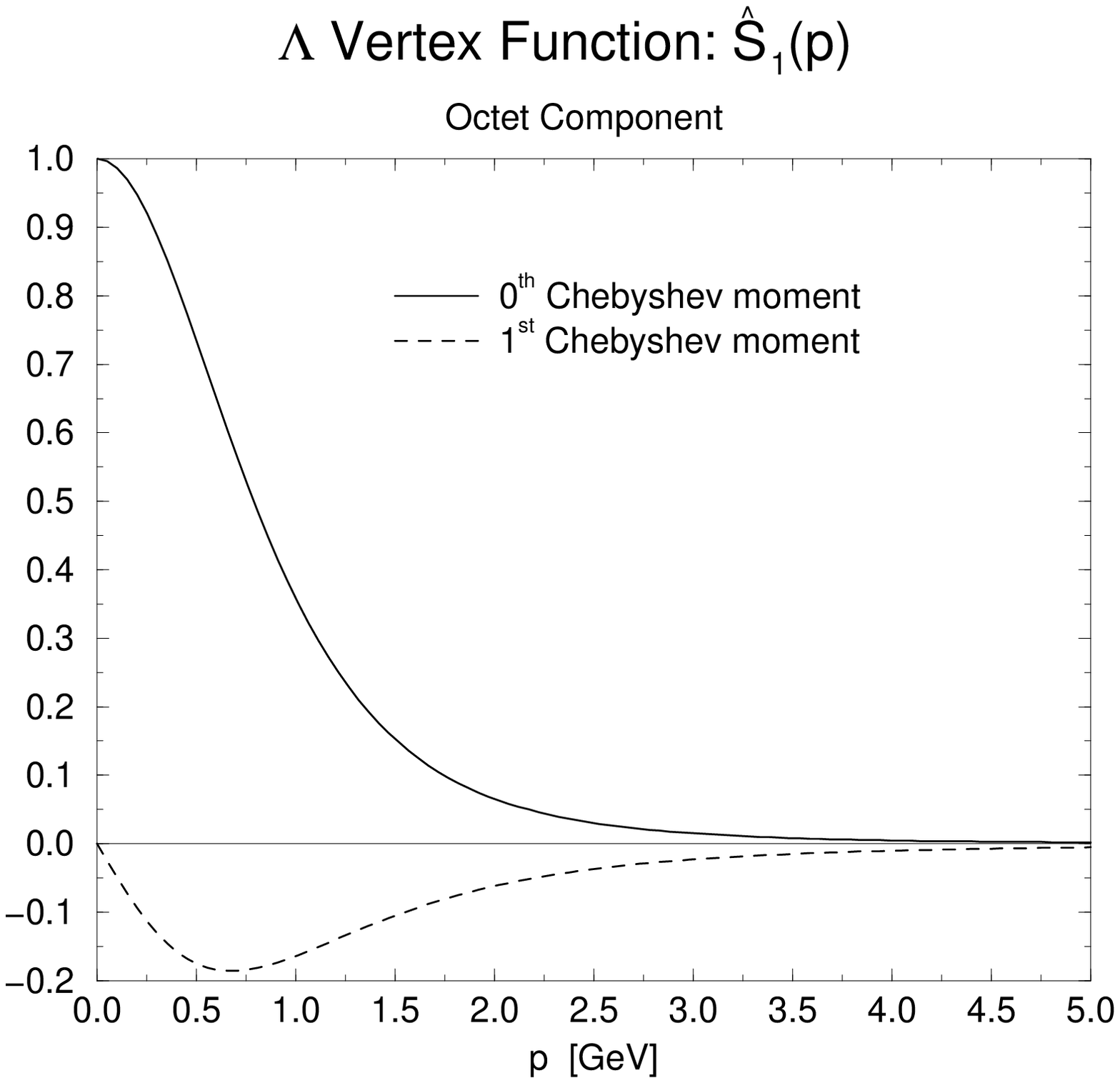}
}{
\epsfxsize 5.9cm
\epsfbox{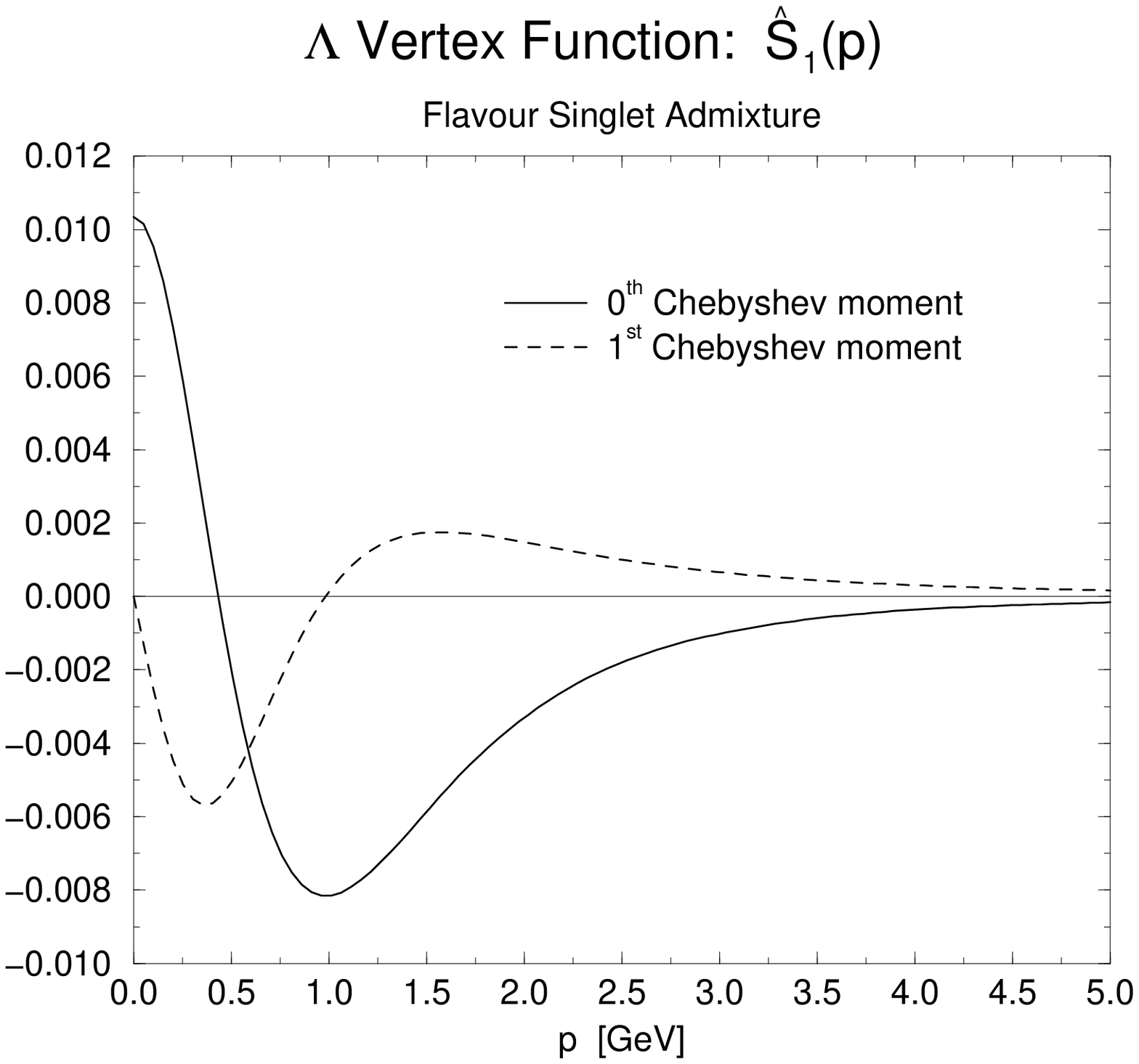}
}}
\centerline{{
\epsfxsize 5.9cm
\epsfbox{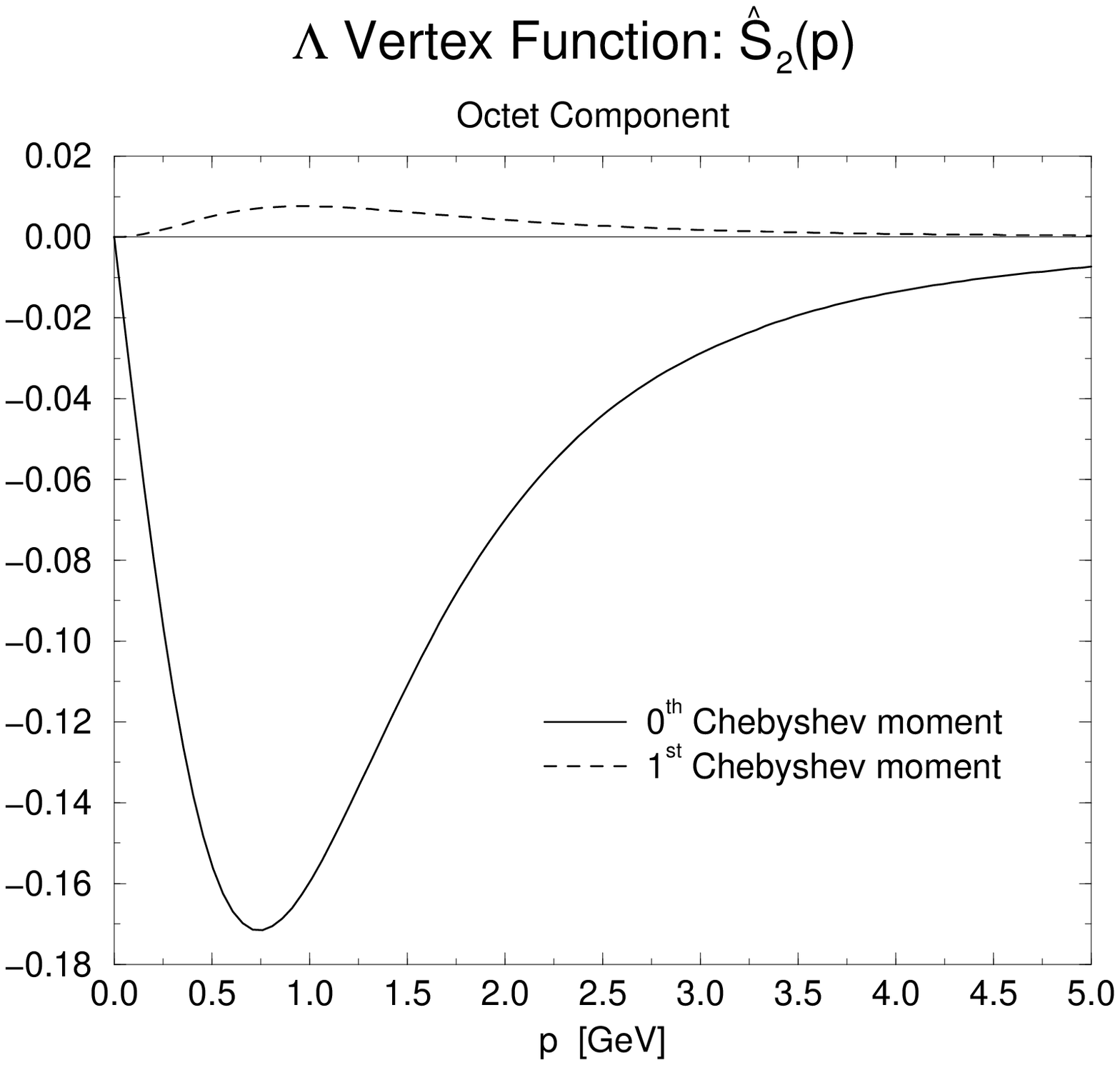}
}{
\epsfxsize 5.9cm
\epsfbox{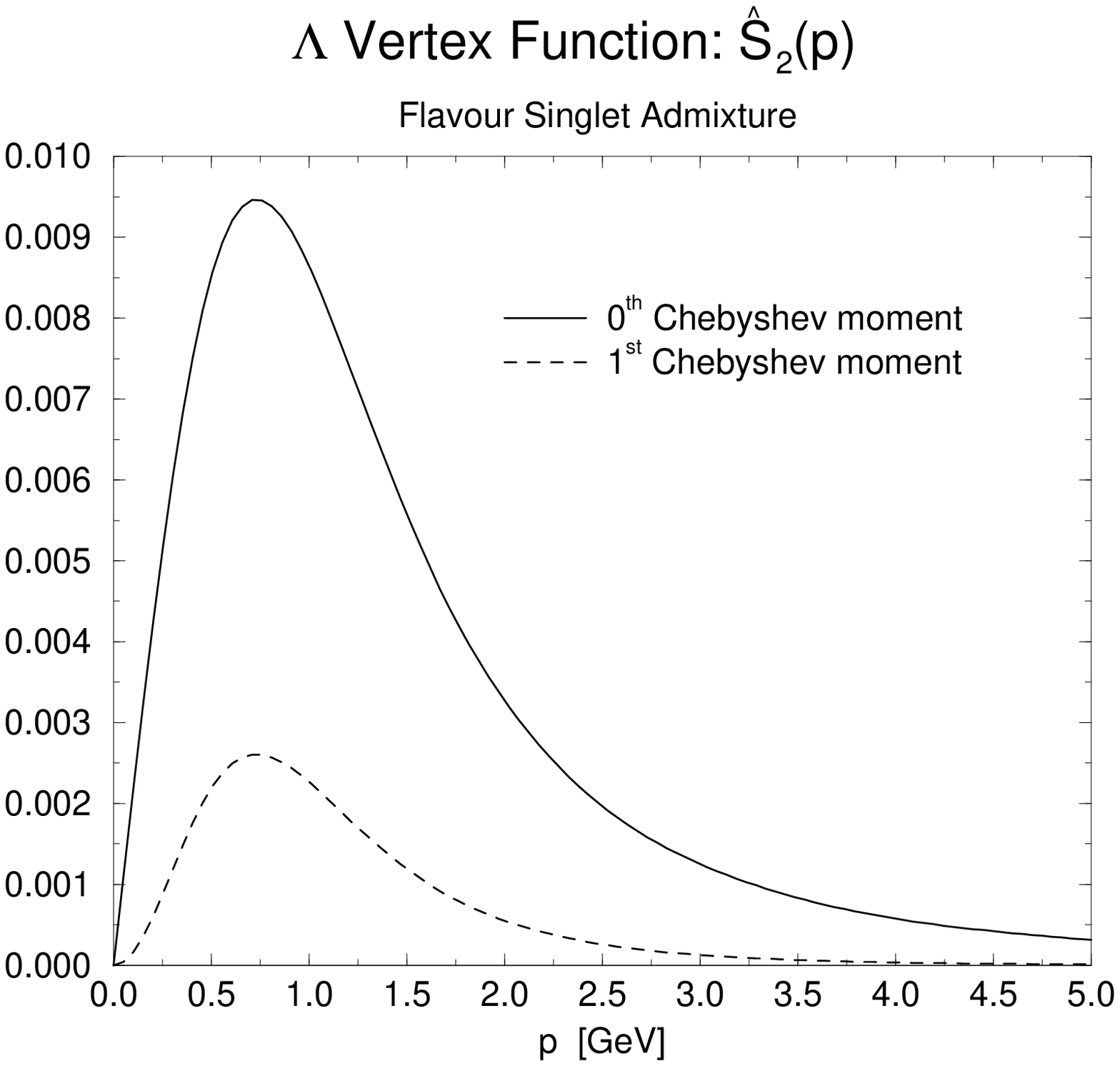}
}}
\caption{\sf
Scalar amplitudes of the vertex function for the $\Lambda$-hyperon,
normalised to $\hat S_{1,octet}^0(p_1)$=1 with $p_1$ being the first point
on the momentum mesh.
The used parameters are \mbox{$m_u$=0.5 GeV, $m_s$=0.65 GeV, $\xi$=1, 
$\eta$=0.33, $d$=10
and $\Lambda$=1 GeV}. \label{slambda}}
\end{figure}
The $\Lambda$-hyperon appears to be of special interest. First, its
measured polarisation asymmetry in the process $p\gamma \rightarrow 
K^+\Lambda$ could provide  a stringent test for the diquark-quark model.
As discussed in \cite{Kro97}, there are only scalar diquarks involved in 
this process and in the following we will concentrate on the scalar
diquark part of the vertex function.

Secondly, broken $SU(3)$-flavour symmetry induces a component
of the total antisymmetric flavour singlet $\frac{1}{\sqrt{3}}
\left[(su)d+(ud)s+(ds)u\right]$\footnote
{With round brackets we denote scalar diquarks, e.g. $(us)=us-su$.
Axialvector diquarks we denote as $[us]=us+su$.}
into wave and vertex function
(see appendix \ref{flavoureq}){\footnote {In nonrelativistic quark 
models with $SU(6)$ symmetry such a component is forbidden 
by the Pauli principle.}}.
As the flavour singlet is only  composed
of scalar diquarks and quarks, this generates 
two additional scalar amplitudes
$\hat S_{1,singlet}$ and $\hat S_{2,singlet}$ besides the usual two from
the octet $\Lambda$ state $\frac{1}{\sqrt{6}}\left[
(su)d-\sqrt{2}(ud)s+(ds)u\right]$. The vector part of the vertex function
remains unchanged in flavour space, $\frac{1}{\sqrt{2}}\left(
[su]^\mu d-[ds]^\mu u\right)$.

These scalar amplitudes for the vertex function
are depicted in figure \ref{slambda} for the 
parameter set of column I in table \ref{baryonmasses}. The
$\hat S_{1,singlet}$-component is suppressed against $\hat S_{1,octet}$ by
two orders of magnitude. However, the purely relativistic $\hat S_{2,singlet}$-
component is only 5 times smaller than its octet counterpart. Observing
that the $\hat S_1$-component usually contributes the major part
to observables as demonstrated in \cite{Hel97}, we can safely
regard the $\Lambda$-hyperon as an almost pure octet state in flavour space.

\pagebreak
\subsection{Wave Functions for Nucleon and Delta} \label{wf}

In this subsection we present Bethe-Salpeter wave 
functions for nucleon 
and delta, using
the parameter set of column I in table \ref{baryonmasses}.
They are normalised to $S_1^0(p_1)$=1 (nucleon) or $D_1^0(p_1)$=1 (delta) where
$p_1$ is the smallest point of the momentum mesh.
The amplitudes represent the strengths of the (${\bf L}^2$, ${\bf S}^2$)
eigenfunctions given in sec. \ref{LSeig} and are simple linear combinations out of the
amplitudes defined in (\ref{NUCGEN},\ref{DELGEN}).

As already mentioned, the convergence of the wave function amplitudes in terms
of Chebyshev polynomials is somewhat slower than for the vertex amplitudes:
Second and zeroth Chebyshev moment of the amplitudes differ by less than
one order of magnitude.
All wave function amplitudes are concentrated to four-momenta
$p \le 0.6$ GeV. 

\begin{figure}[t]
\centerline{
\epsfxsize 5.9cm
\epsfbox{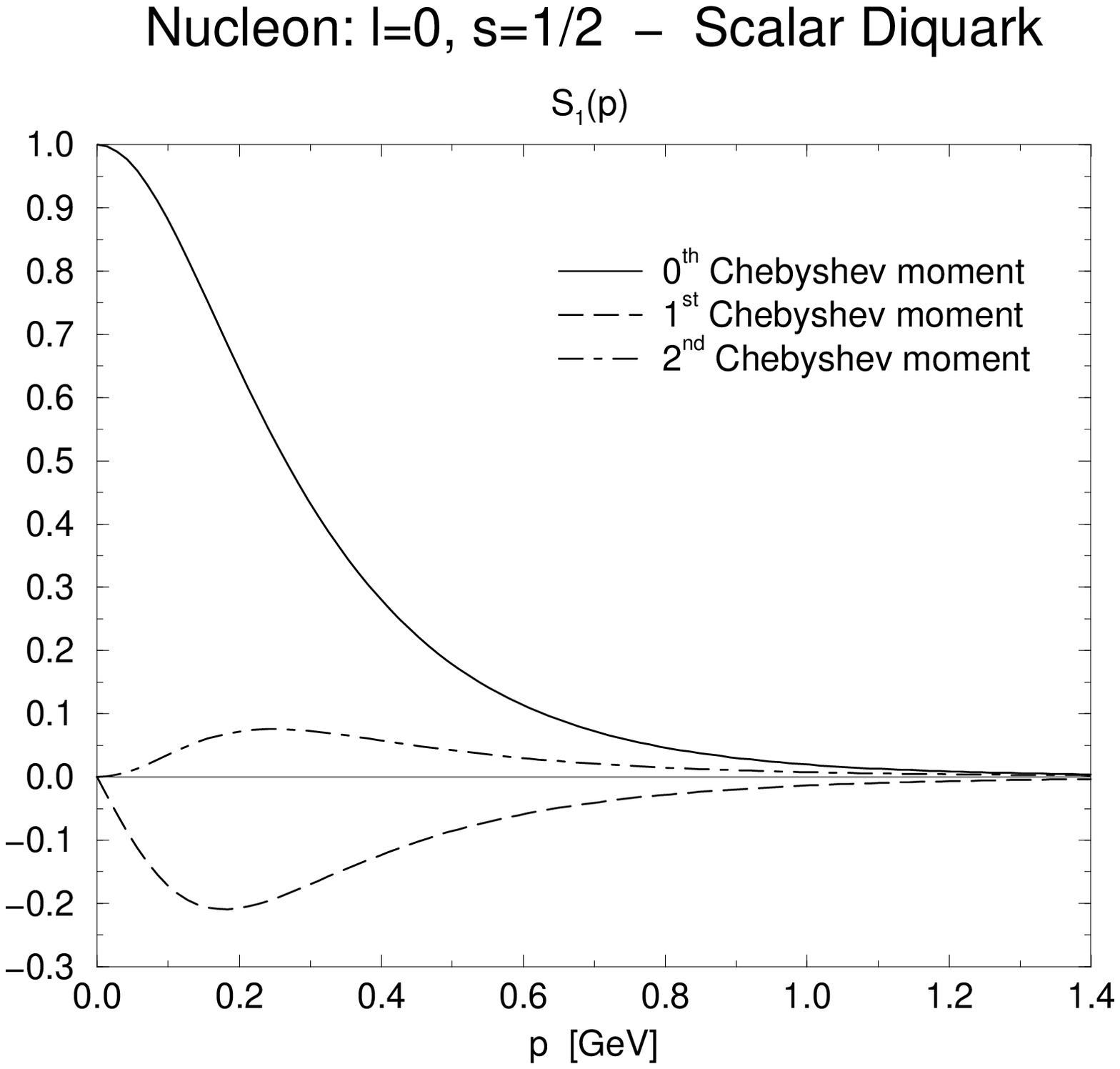}
\epsfxsize 5.9cm
\epsfbox{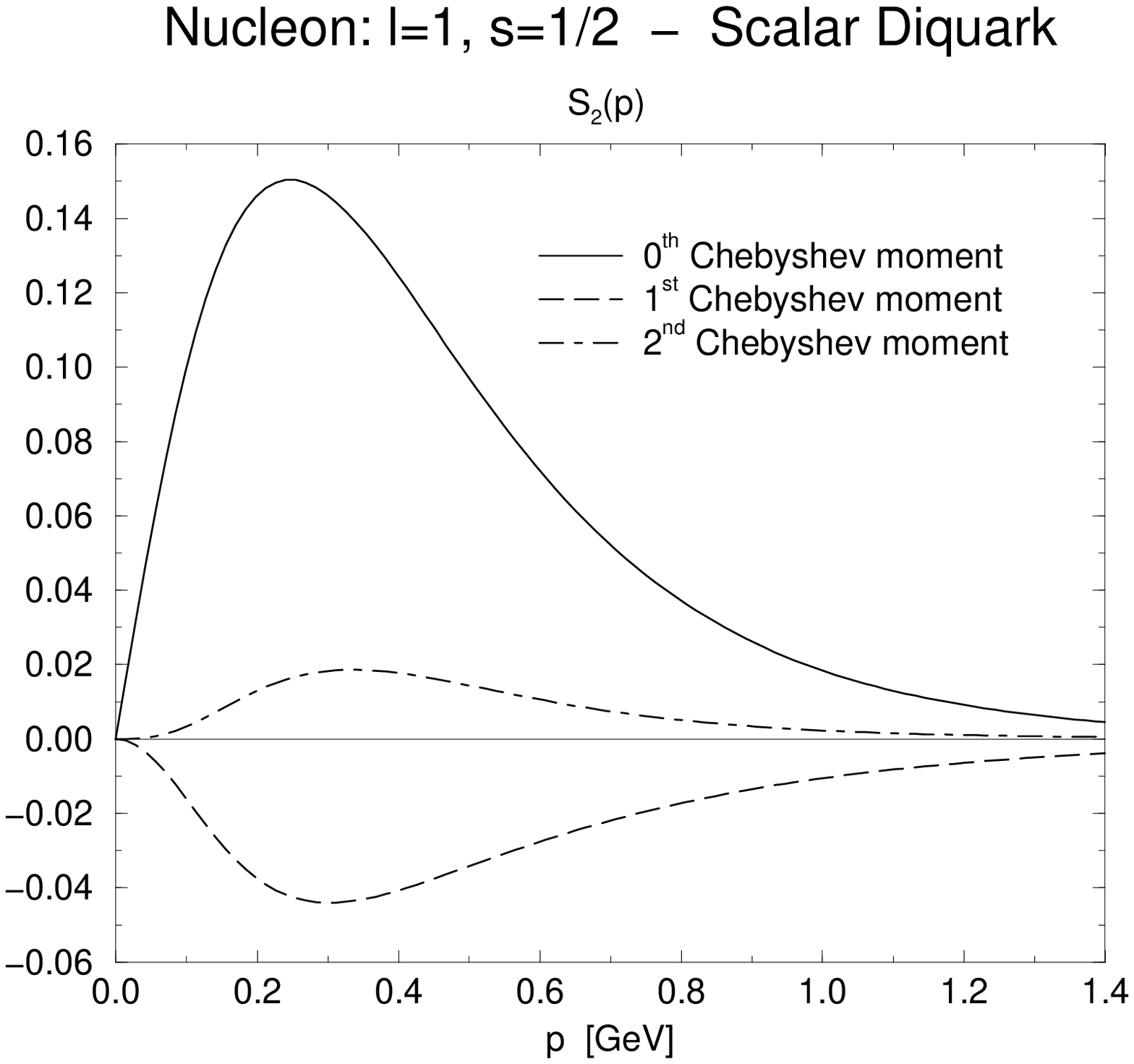}
}
\centerline{
\epsfxsize 5.9cm
\epsfbox{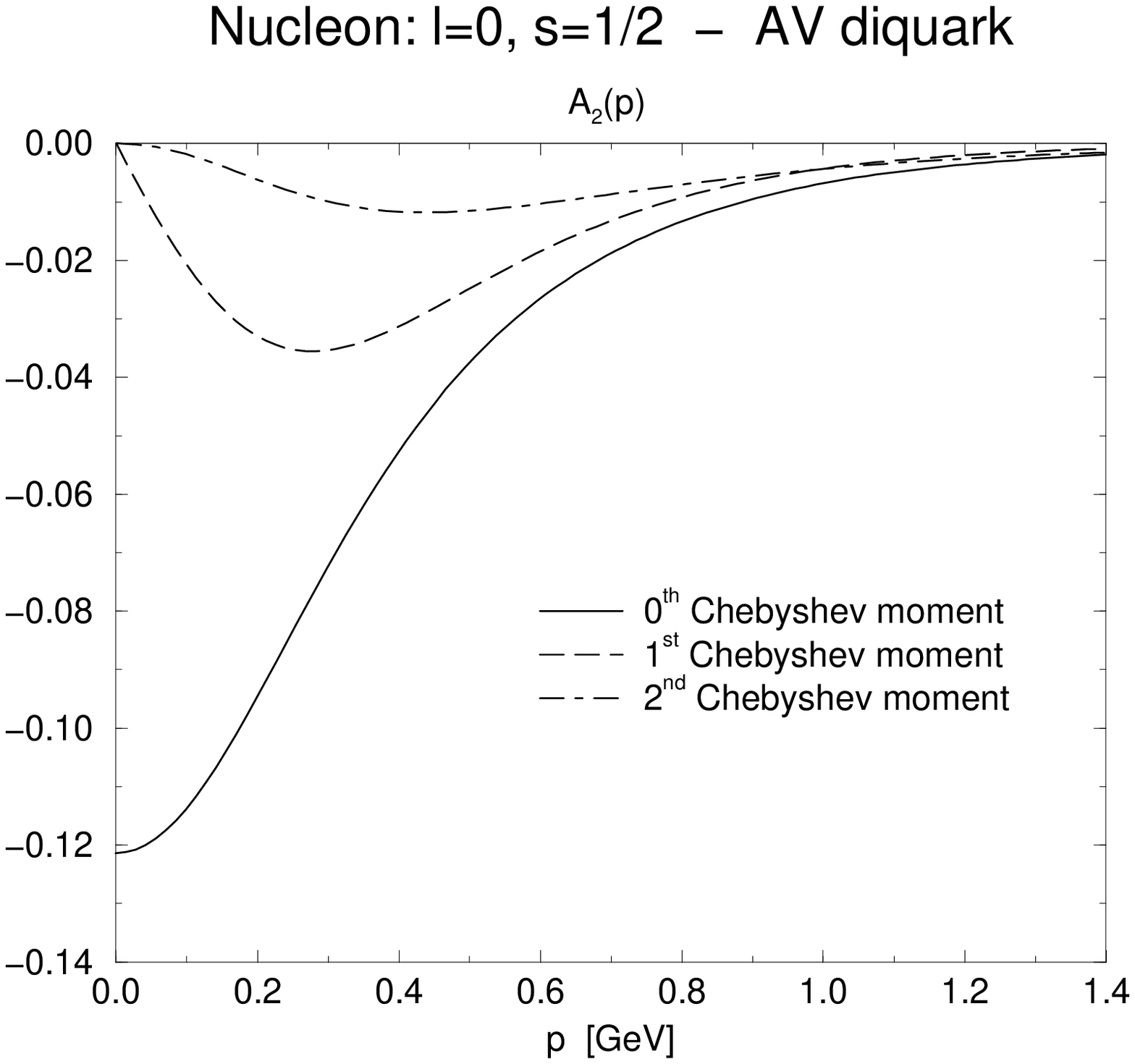}
\epsfxsize 5.9cm
\epsfbox{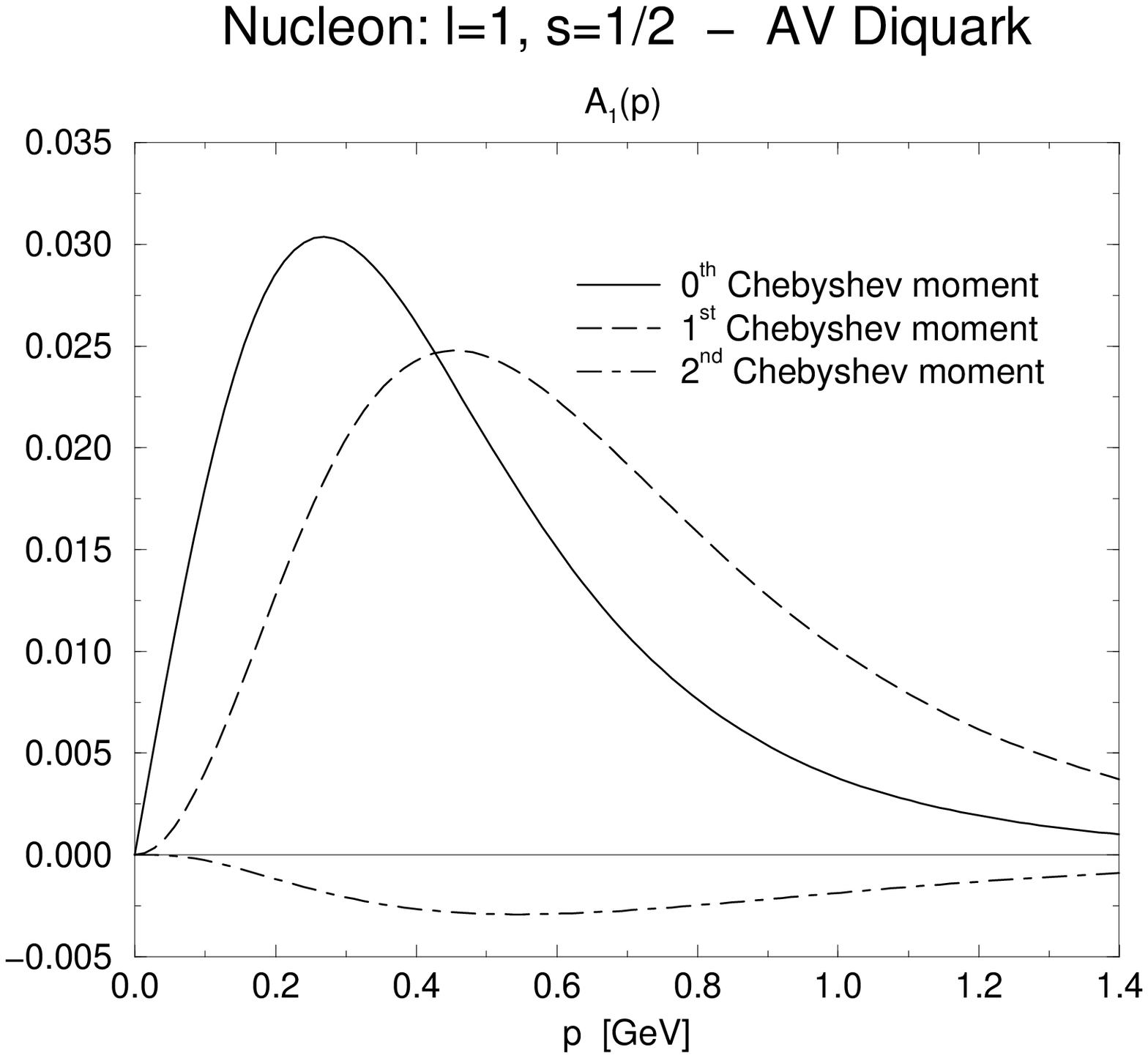}
}
\centerline{
\epsfxsize 5.9cm
\epsfbox{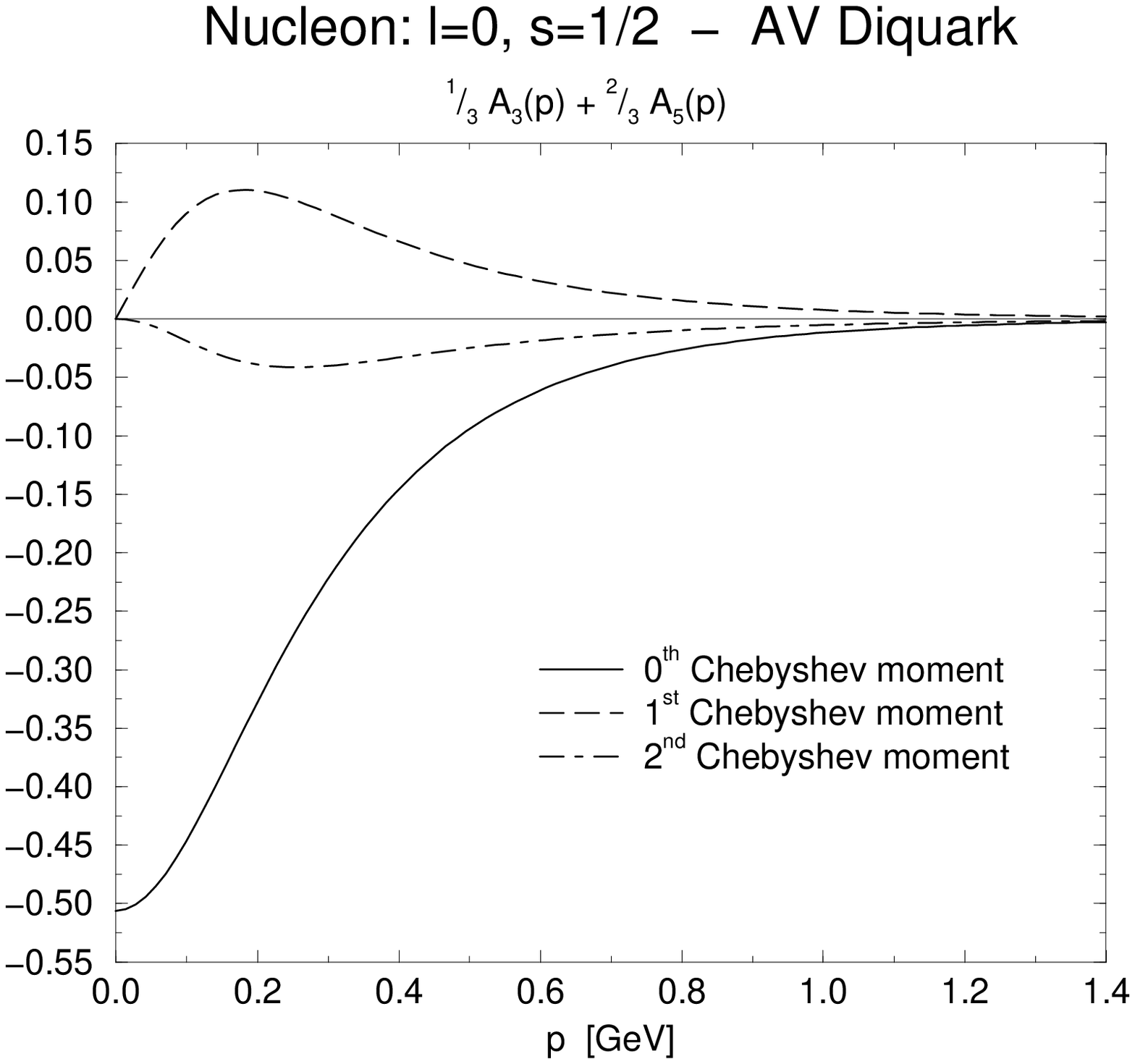}
\epsfxsize 5.9cm
\epsfbox{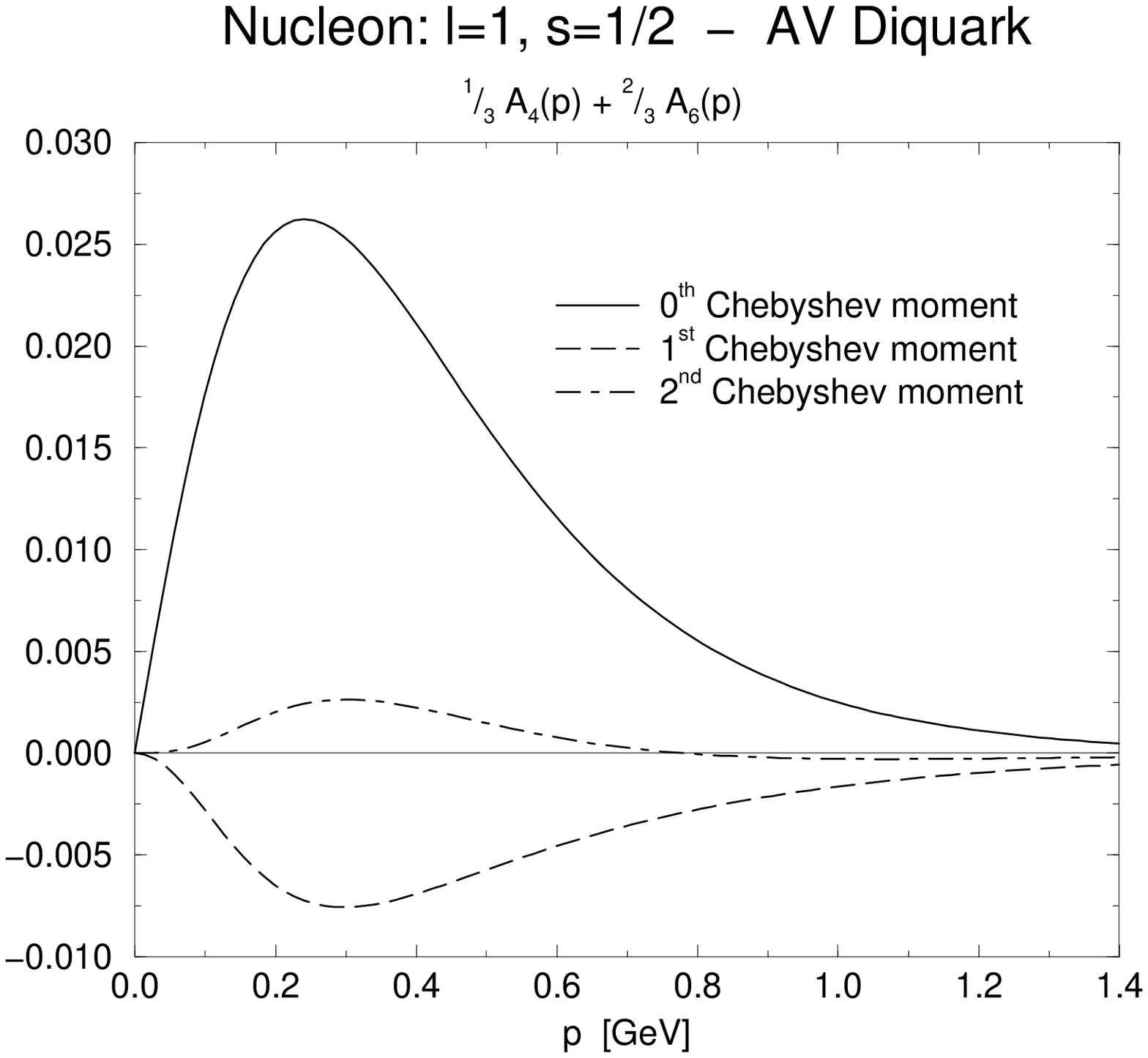}
}
\centerline{
\epsfxsize 5.9cm
\epsfbox{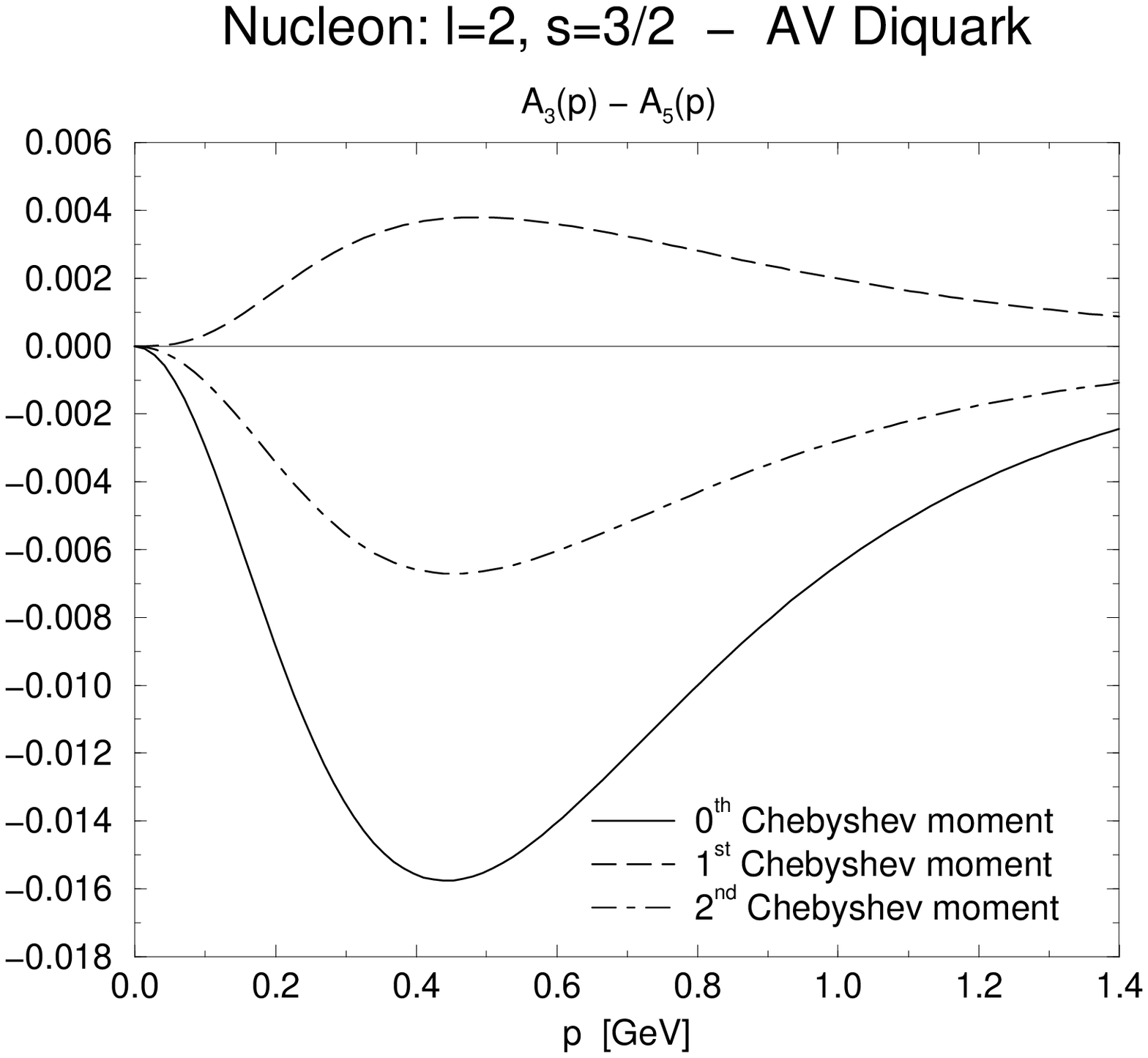}
\epsfxsize 5.9cm
\epsfbox{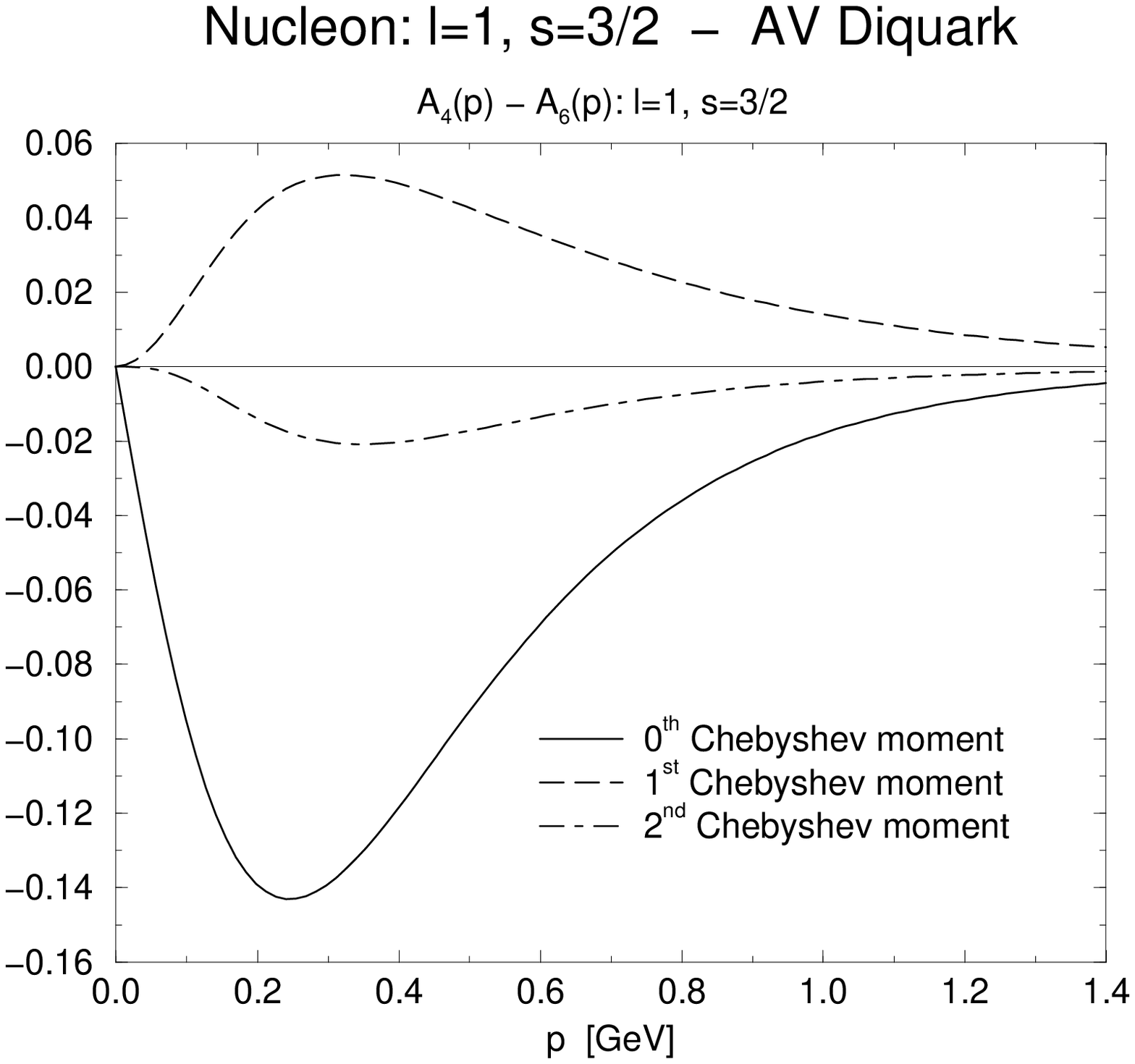}
}
\caption{\sf
Scalar and axial vector amplitudes of the nucleon wave function with
parameters given by $m_u$=0.5 GeV, $\xi$=1, $\eta$=0.33, $d$=10 and
$\Lambda$=1 GeV. \label{wfnuc}}
\end{figure}

In figure \ref{wfnuc}, the nucleon amplitudes with even
orbital angular momentum $l$ appear in the left row: These are the three
$s$-waves describing (i) scalar diquark and quark, (ii,iii)
axialvector diquark and quark oppositely aligned to give spin 1/2.
There are two axialvector diquark components due to the virtual time
component of the latter. The scalar diquark component is the most important
but the other $s$-waves enhance binding by approximately 30 \%. The fourth 
``non-relativistic'' component is a strongly 
suppressed $d$-wave with quark and axialvector diquark aligned to give spin 3/2.
 The lower components depicted on the right side can be
understood as the admixture of negative-energy spinors to the proton
wave function and contribute approximately 10 \% to the binding energy.

\begin{figure}[t]
\centerline{
\epsfxsize 5.9cm
\epsfbox{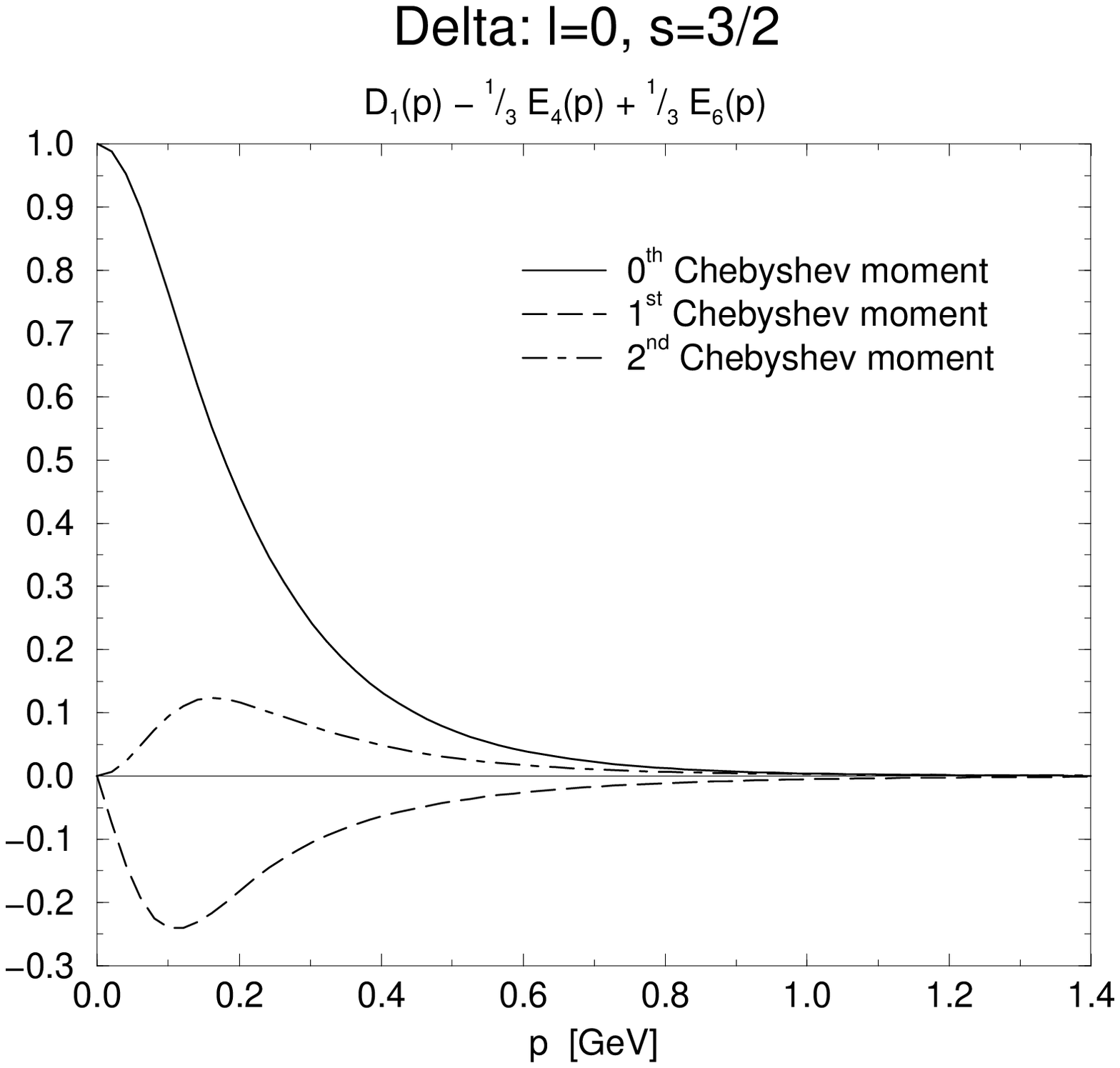}
\epsfxsize 5.9cm
\epsfbox{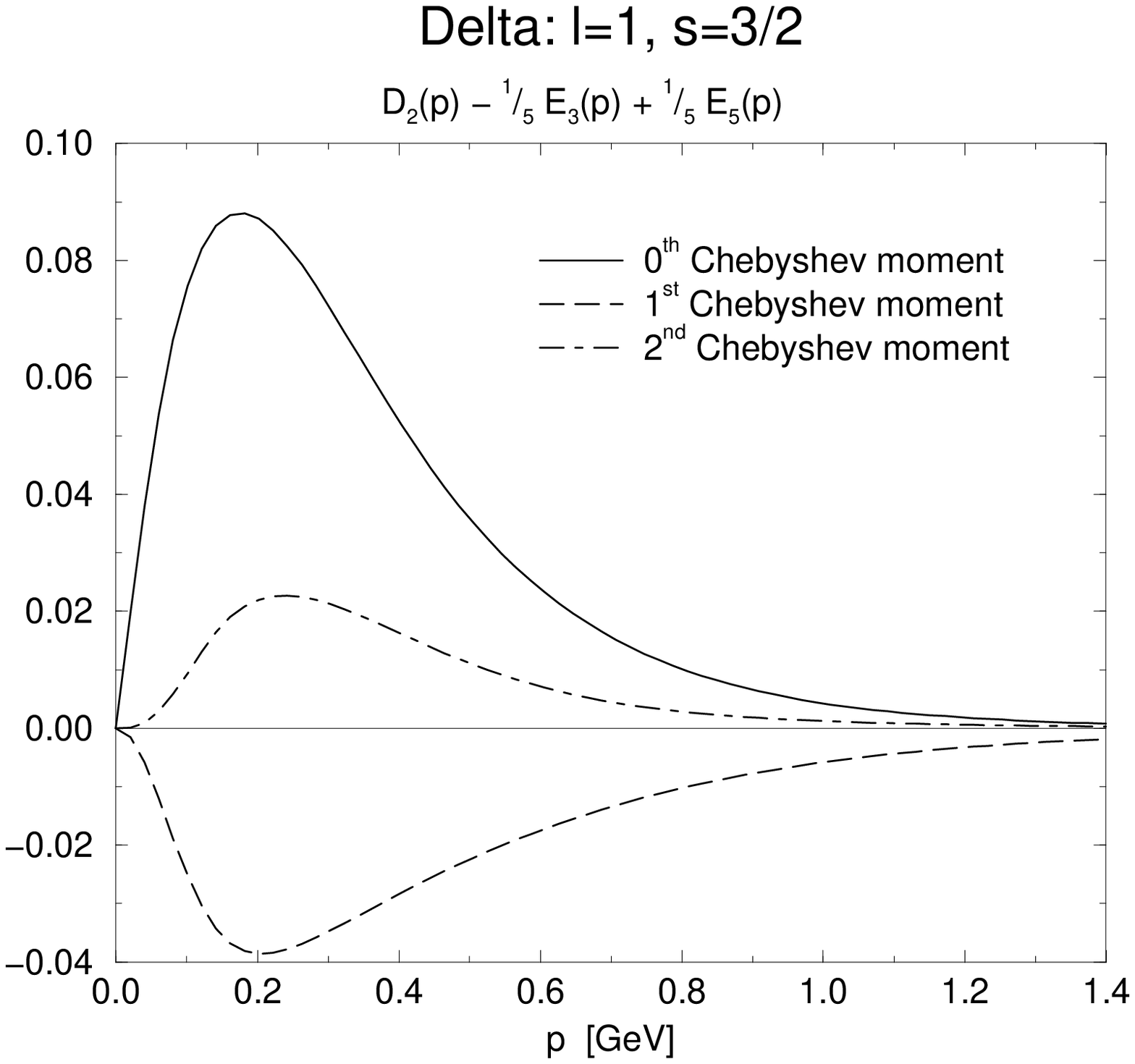}
}
\centerline{
\epsfxsize 5.9cm
\epsfbox{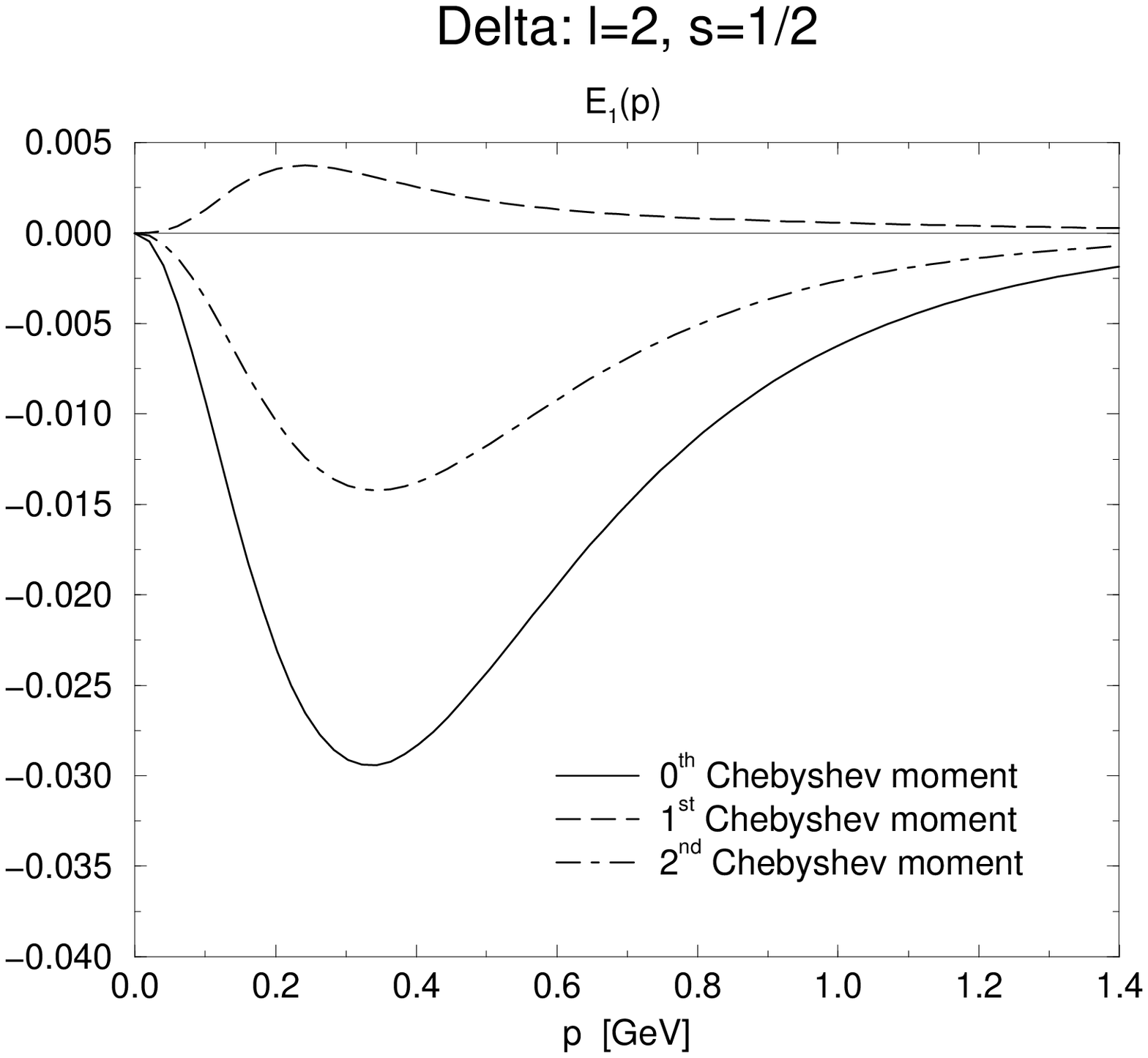}
\epsfxsize 5.9cm
\epsfbox{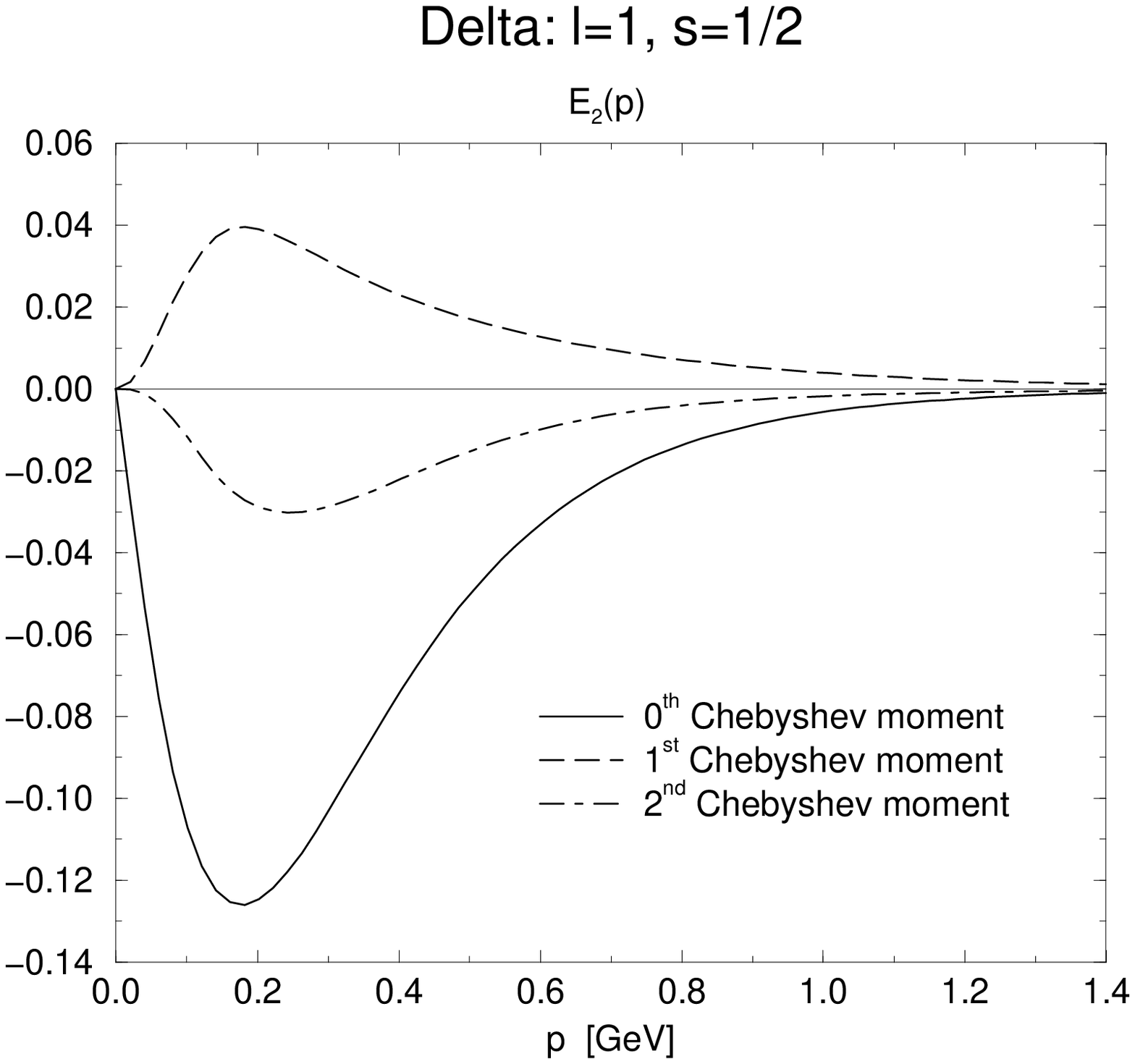}
}
\centerline{
\epsfxsize 5.9cm
\epsfbox{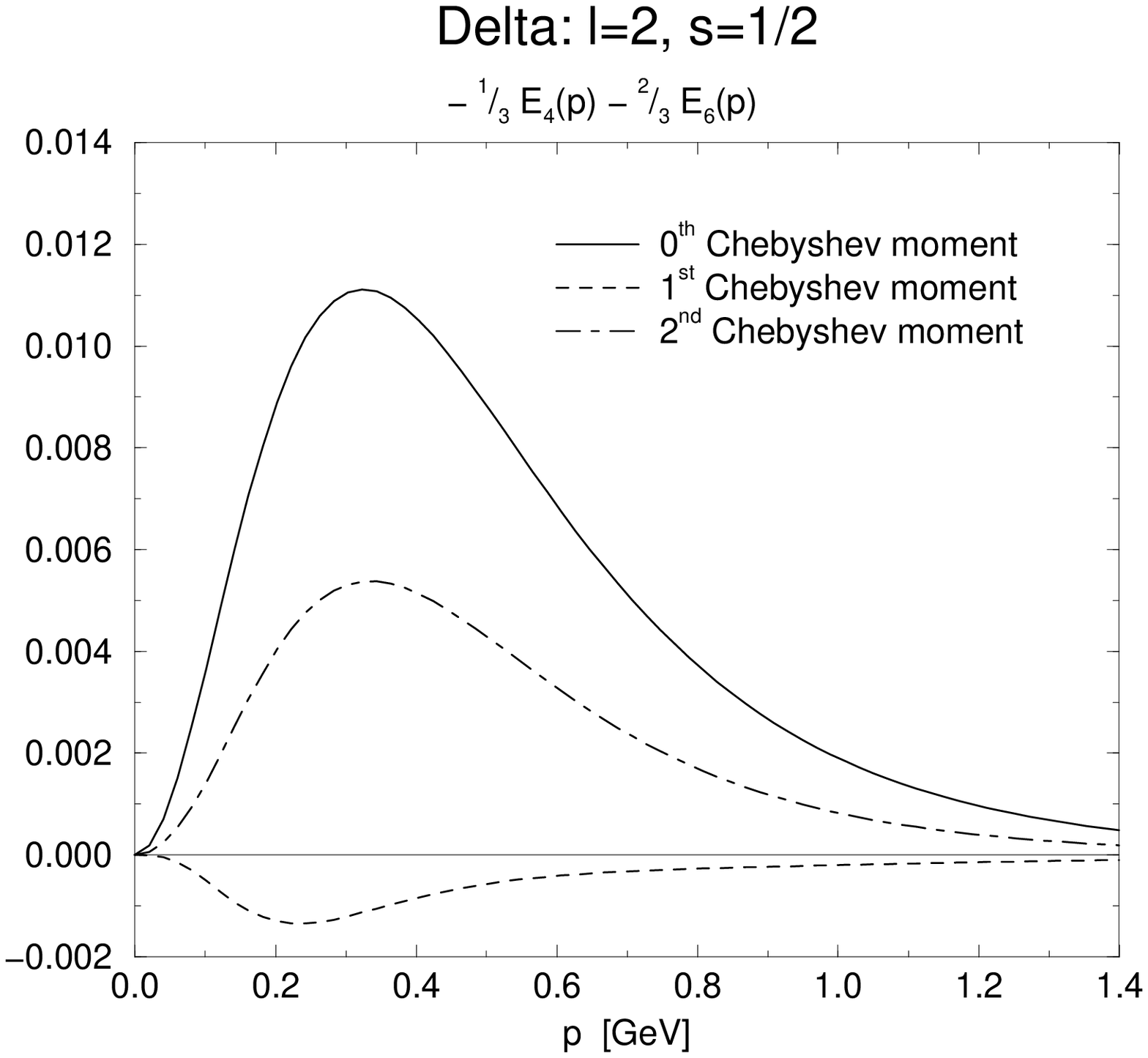}
\epsfxsize 5.9cm
\epsfbox{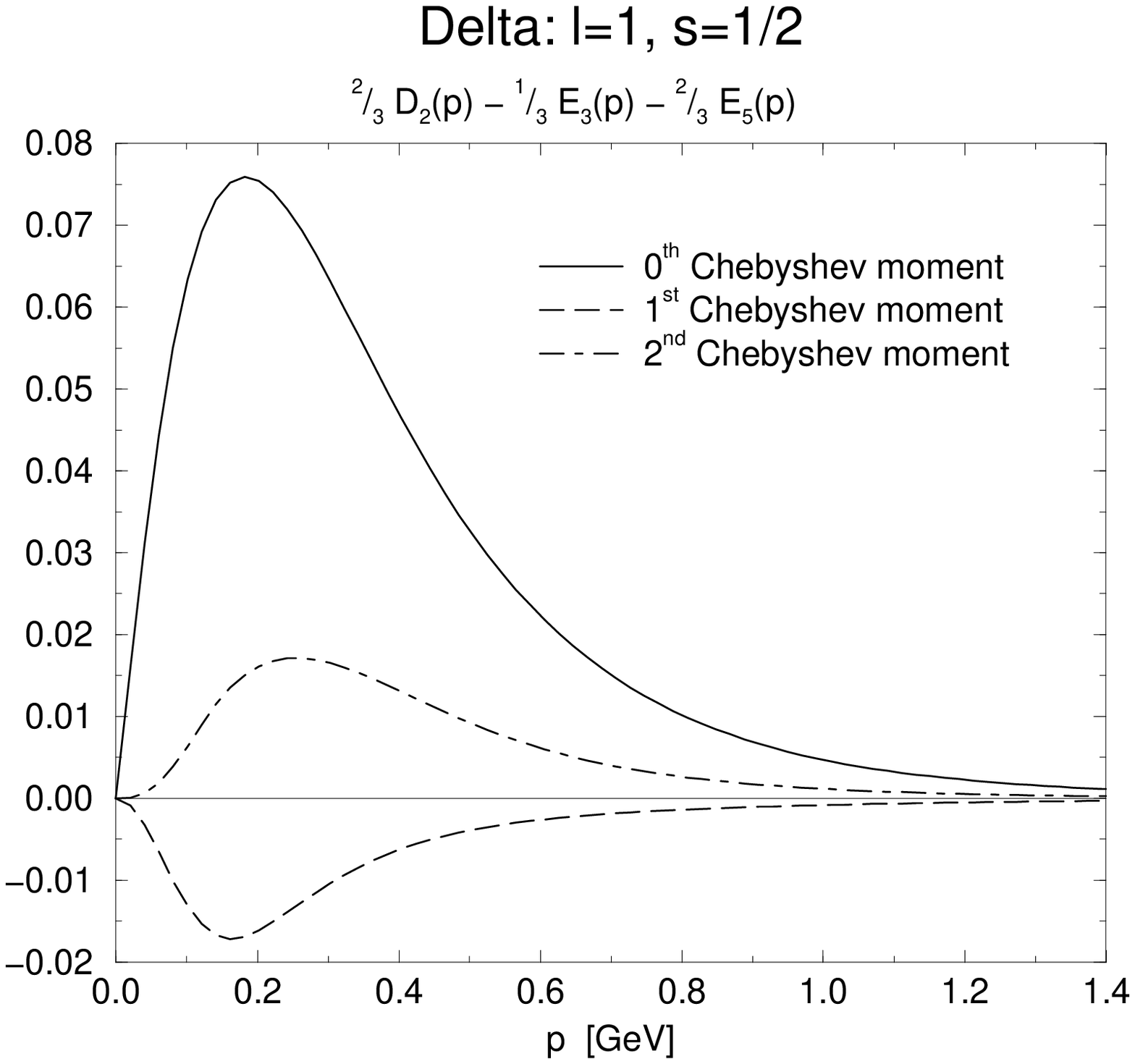}
}
\centerline{
\epsfxsize 5.9cm
\epsfbox{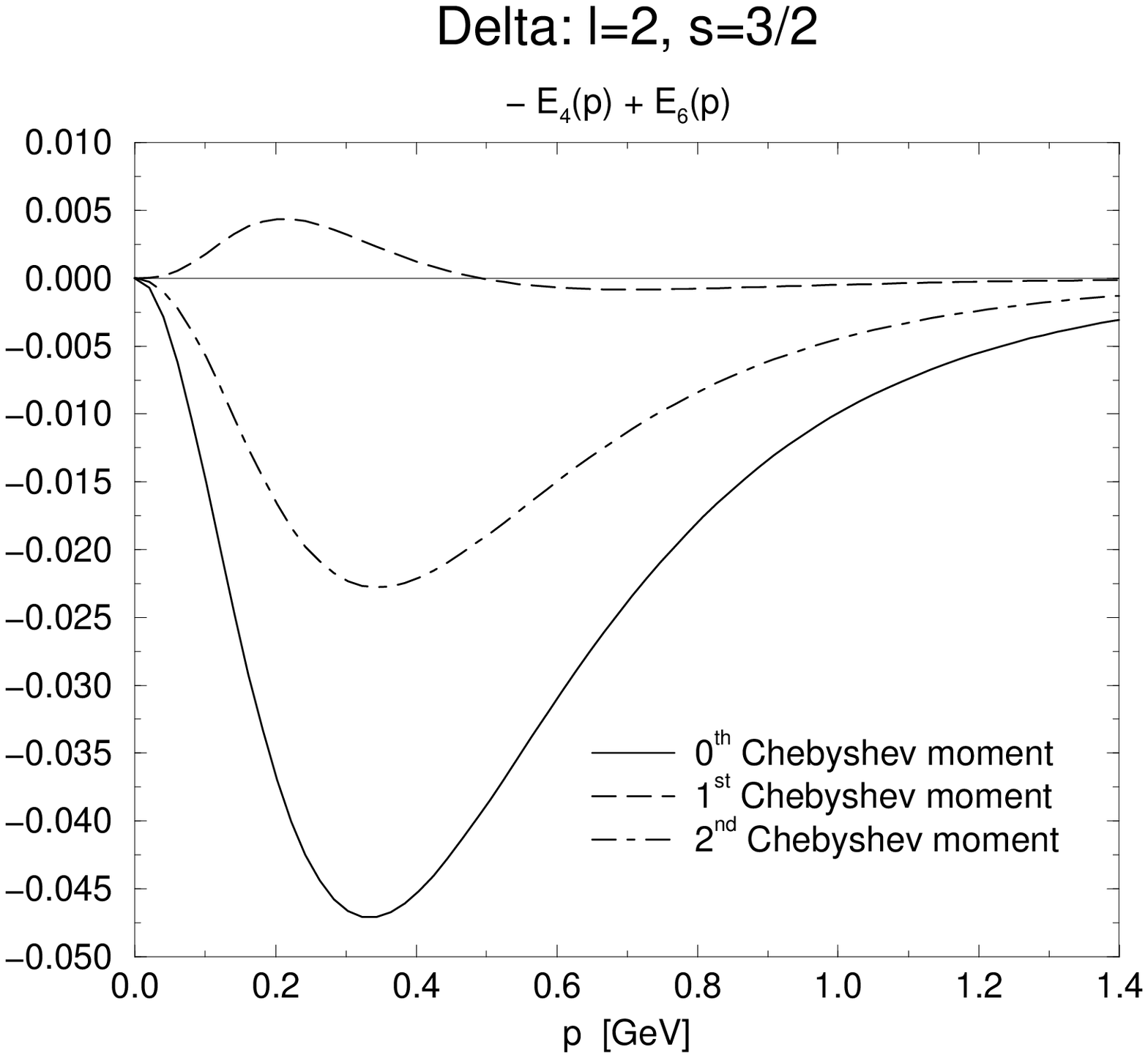}
\epsfxsize 5.9cm
\epsfbox{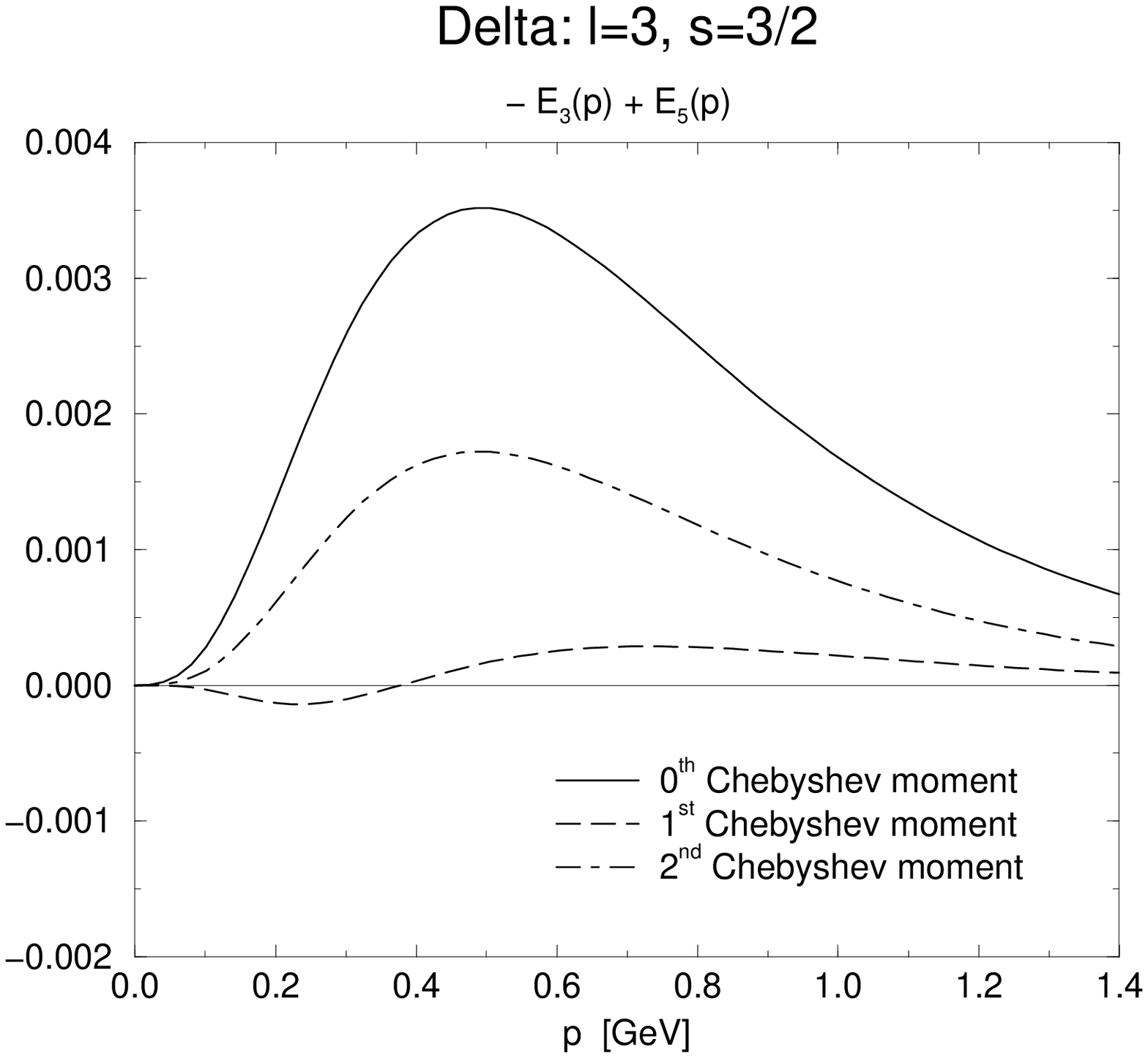}
}
\caption{\sf
Axial vector amplitudes of the delta wave function with
parameters given by $m_u$=0.5 GeV, $\xi$=1, $\eta$=0.33, $d$=10 and
$\Lambda$=1 GeV. \label{wfdel}}
\end{figure}

The delta amplitudes in fig.\ref{wfdel} have also been arranged
into ``non-relativistic'' (left row) and ``relativistic'' components (right row).
As expected, the only $s$-wave dominates the decomposition,
but the relativistic $p$-wave components, which act repulsively, increase
the eigenvalue by approximately 20 \% and are thus non-negligible. Again
the component with the highest orbital angular momentum, $l$=3, is
highly suppressed.

No clear effects of flavour symmetry breaking, as in terms of the width in 
momentum space, can be detected for the other octet wave functions when 
compared to the nucleon. For the vertex functions (where the zeroth Chebyshev
moment dominates), we observed a limited increase of the width in momentum
space with increasing number of $s$-quarks (10\% difference between $N$
and $\Xi$). The width is hereby defined as the absolute value
of the relative momentum $p$ where the zeroth Chebyshev moment of an $s$-wave
amplitude reaches half the maximum value.

\clearpage

\pagebreak[4]
\clearpage

\section{Conclusions and Outlook}

In this paper we extended the covariant and confining
diquark-quark model of ref. \cite{Hel97}
by including axialvector diquarks in the description 
of baryons within the Bethe-Salpeter approach. 
Thus we were able to calculate the octet
baryons masses as well as the decuplet masses and wave functions. 
We implemented confinement via an effective parametrisation of
the constituent propagators and demonstrated the existence of 
bound states beyond the pseudo-threshold.

We decomposed octet and decuplet vertex and wave 
functions in the Dirac and Lorentz algebra, obtaining
8 scalar functions, respectively, which we computed 
numerically. Two approximations to  the Bethe-Salpeter equation were
discussed and compared: The direct ladder approximation,
and a modified ladder approximation 
with a special momentum routing leading to a kernel 
independent of the bound state mass.

In order to fix the parameters of the model preliminarily, we
computed the masses of octet and decuplet for
broken $SU(3)_{flavour}$ with isopin conserved. With the scalar and
axialvector diquark mass assumed to be equal, the octet-decuplet mass 
splitting is a result of the different effects of the coupling
constants  in the scalar
and axialvector diquark channel.
For a parameter set which fits the octet and decuplet masses
well, we computed  vertex amplitudes and wave functions for 
all octet and decuplet baryons. The wave functions for baryons with different
strangeness content but same spin differ mostly due to the different
flavour Clebsch-Gordan coefficients, the respective scalar functions being
very similar. Therefore
we presented only the wave functions for the 
nucleon and delta out of this data. 
The decomposition of the wave functions in the rest frame of the
bound state in terms of spin and orbital angular momentum eigenstates 
revealed an $s$-wave dominance in all ground state baryons stemming from both scalar 
and axialvector diquark contributions. The 
$p$-wave contributions sneaking in via the
lower components of the spinors are of greater importance for decuplet baryons
than for octet baryons.

Due to its special
role among the other baryons, we investigated the  
$\Lambda$-hyperon in more detail and discussed its vertex amplitudes.
In our approach, the $\Lambda$ acquires a small flavour singlet 
admixture which is absent in $SU(6)$ symmetric 
nonrelativistic quark models.

This work, together with our preceding paper \cite{Hel97},
provides a sound basis for further applications of this 
approach to baryon phenomenology. 
The calculated amplitudes which encode the non-trivial 
information of a baryon as a diquark-quark bound state
serve  as a necessary  input for the calculation of 
various observables. A calculation of  the electromagnetic
form factors  of the octet and decuplet baryons is hereby
the next task. Such an investigation will furthermore help to
fix some of the parameters. Our aim is however to apply the
covariant and confining diquark-quark model to processes, which are far less
understood. Especially, 
the reactions   
$p\gamma\rightarrow K\Lambda$ and
$pp\rightarrow p \Lambda K$, currently measured at ELSA and COSY,
respectively, will serve as a stringent test of our 
approach.  
Additionally, we plan to get further insight into this picture of baryons
by computing structure functions for a spectator model \cite{Kus97}
which includes the axialvector diquark.

\vspace{0.2cm}
\noindent
{\bf Acknowledgement}\\
We thank Lorenz v. Smekal and Herbert Weigel for valuable discussions,
especially the latter for insisting on his critical remarks.
We are grateful to Lorenz v. Smekal for the possibility to compare with
his numerical results regarding the Bethe-Salpeter eigenvalues prior to
publication.

\appendix

\section{Octet and Decuplet Equations} \label{flavoureq}
The symmetric and antisymmetric flavour matrices can be written down as:
{\small
\begin{eqnarray}
 t^a_{\mathcal A} & = & \{ \rho^{a=1\dots3} \}= \nonumber \\
                  & = &  \left\{ \pmatrix{0 & 1 & 0 \cr 
        -1 & 0 & 0 \cr 0 & 0 & 0\cr},
        \pmatrix{0 & 0 & 1 \cr 0 & 0 & 0 
        \cr -1 & 0 & 0\cr},
        \pmatrix{0 & 0 & 0 \cr 0 & 0 & 1 
        \cr 0 & -1 & 0\cr} \right\}, \label{lambdadef} \\
 t^a_{\mathcal S} & = & \{ \rho^{a=4\dots9} \}= \nonumber \\
                  & = & \left\{ \pmatrix{\sqrt{2} & 0 & 0 \cr
        0 & 0 & 0 \cr 0 & 0 & 0\cr},
  \pmatrix{0 & 0 & 0 \cr 0 & \sqrt{2} & 0 \cr 0 &0 & 0\cr},
  \pmatrix{0 & 0 & 0 \cr 0 & 0 & 0 
  \cr 0 & 0 & \sqrt{2}\cr}\right., \nonumber\\
                  &   & \left. \pmatrix{0 & 1 & 0 
  \cr 1 & 0 & 0 \cr 0 & 0 & 0\cr},
  \pmatrix{0 & 0 & 1 \cr 0 & 0 & 0 \cr 1 & 0 & 0\cr},
  \pmatrix{0 & 0 & 0 \cr 0 & 0 & 1 
  \cr 0 & 1 & 0} \right\}. 
\end{eqnarray}}
By these conventions flavour antisymmetric diquarks are $(ud):=ud-du,$ etc.
and flavour symmetric diquarks $\sqrt{2}uu,[ud]:=ud+du,$ etc. The flavour wave 
functions of octet and decuplet states 
do not decouple  once the
$s$-quark breaks the symmetry. The Bethe-Salpeter equation (\ref{sc})
still describes nucleons (isospin is assumed to be conserved), 
and equation (\ref{deltaBSE}) still refers to $\Delta$ and $\Omega$ which
possess only single-component flavour wave functions $[uu]^{\mu\rho} u$ resp.
$[ss]^{\mu\rho} s$.

We use the following abbreviations to give short-hand Bethe-Salpeter equations
for the remaining baryons:\\
\vspace{0.3cm}
\begin{tabular}{|lcl|} \hline
$(ab)c$ resp. $[ab]^\mu c$ &:& octet flavour wave functions with diquark 
                               flavour  \\
                           & &  content $ab$ and spectator quark of flavour $c$ \\
$[ab]^{\mu\rho} c$ &:& decuplet flavour wave functions  \\                             
$S_a$            &:& quark propagator of the spectator quark $a$ \\
$D_{(ab)}$ resp. $D^{\mu\mu'}_{[ab]}$ &:& scalar resp. axialvector diquark 
                            propagator\\
$K_a^{\rho\lambda}$ = $\int\frac{d^4p^{\prime}}{(2\pi)^4}
\gamma^{\rho} \tilde S_a(-q) \gamma^{\lambda}$            
&:& exchange kernel for quark flavour $a$, see eq.(\ref{dpff}) for the 

\\
\hspace*{0.5cm}{\small($\rho,\lambda=1\dots 5)$}& & definition of $\tilde S$\\ \hline                        
\end{tabular}\\
\vspace{0.3cm}
Of course, different to eqs. (\ref{S}-\ref{Da}), different masses
corresponding to the flavour content of quark and diquark are used
in the numerators and denominators of the propagators.
  
For octet states
each flavour wave function with scalar diquarks is to be expanded
in Dirac space according to eq. (\ref{expnuc}) with only the scalar
amplitudes $S_1$ and $S_2$ and each one with axialvector diquarks
according to the same equation with the 6 vector amplitudes $A_1\dots A_6$.
For decuplet states only flavour wave functions with axialvector 
diquarks are considered,
they have to be expanded as indicated in eq. (\ref{expdel}).

The Bethe-Salpeter equation for $\Xi$ baryons now
reads:

\begin{eqnarray} \label{xi}
\pmatrix{(us)s \cr [us]^\mu s \cr  [ss]^\mu u \cr}&=& -g_s^2
\pmatrix{S_s D_{(us)} &0&0\cr0&S_s D^{\mu\mu'}_{[us]}&0\cr
         0&0& S_u D^{\mu\mu'}_{[ss]}\cr}  \\
& & \times
\pmatrix{K_u^{5,5} & -\frac{g_a}{g_s}K^{\nu,5}_u & 
             \sqrt{2}\frac{g_a}{g_s}K^{\nu,5}_s \cr
         -\frac{g_a}{g_s}K^{5,\mu'}_u & \frac{g_a^2}{g_s^2}K^{\nu\mu'}_u &
          \sqrt{2}\frac{g_a^2}{g_s^2}K^{\nu\mu'}_s \cr
         \sqrt{2}\frac{g_a}{g_s}K^{5,\mu'}_s & 
         \sqrt{2}\frac{g_a^2}{g_s^2}K^{\nu\mu'}_s& 0\cr}
\pmatrix{(us)s \cr [us]^\nu s \cr  [ss]^\nu u \cr} \nonumber
\end{eqnarray}

By interchanging $s \leftrightarrow u$ one obtains immediately the equation
for $\Sigma$ baryons.

Broken $SU(3)$ couples the symmetric $\Lambda$ and the flavour singlet.
We introduce the flavour wave functions $F_1=\frac{1}{\sqrt{2}}
((us)d-(ds)u)$, $F_2=(ud)s$ and $\Lambda^\mu=\frac{1}{\sqrt{2}}
([us]^\mu d-[ds]^\mu u)$ and the equation for the physical $\Lambda$ reads:

\begin{eqnarray}
\pmatrix{F_1 \cr F_2 \cr  \Lambda^\mu \cr} &=& -g_s^2
\pmatrix{S_u D_{(us)} &0&0\cr0&S_s D_{(ud)}&0\cr
         0&0& S_u D^{\mu\mu'}_{[us]}\cr}  \\
& & \times
\pmatrix{-K_s^{5,5} & \sqrt{2}K_u^{5,5} & -\frac{g_a}{g_s}K^{\nu,5}_s \cr
         \sqrt{2}K_u^{5,5} & 0 & -\sqrt{2}\frac{g_a}{g_s}K^{\nu,5}_u \cr
      -\frac{g_a}{g_s}K^{5,\mu'}_s & -\sqrt{2}\frac{g_a}{g_s}K^{5,\mu'}_u& 
      -\frac{g_a^2}{g_s^2}K^{\nu\mu'}_s\cr}
\pmatrix{(us)s \cr [us]^\nu s \cr  [ss]^\nu u \cr} \nonumber
\end{eqnarray}

The $\Xi^*$ baryons belonging to the decuplet are described by the equation:

\begin{equation}
\pmatrix{[us]^{\mu\rho} s \cr  [ss]^{\mu\rho} u \cr}= -g_a^2
\pmatrix{S_s D^{\mu\mu'}_{[us]}&0\cr
         0& S_u D^{\mu\mu'}_{[ss]}\cr}
\pmatrix{K^{\nu\mu'}_u & \sqrt{2} K^{\nu\mu'}_s \cr
         \sqrt{2} K^{\nu\mu'}_s& 0\cr}
\pmatrix{[us]^{\nu\rho} s \cr  [ss]^{\nu\rho} u \cr}
\end{equation}

As before the equation for the $\Sigma^*$ baryons may be obtained
by interchanging $s \leftrightarrow u$. Note that we neglected
contributions from a state with scalar diquark $(us)$.

\section{Solutions with Proca Propagator for Axialvector Diquarks} 
\label{Procadis}

In order to study the influence of the full Proca propagator on 
vertex functions of spin-1/2 baryons and the corresponding
eigenvalues we are enforced to use a diquark
size factor of the dipole type, writing instead of eq. (\ref{dpff})
\be
S(q) \rightarrow \tilde S(q) =
S(q)\left( \frac{\Lambda^2}{q^2+\Lambda^2}\right)^2.
\ee
to regularise the equation
which has no stable solution otherwise.

\begin{figure}[t]
\centerline{\epsfxsize 7.5cm
\epsfbox{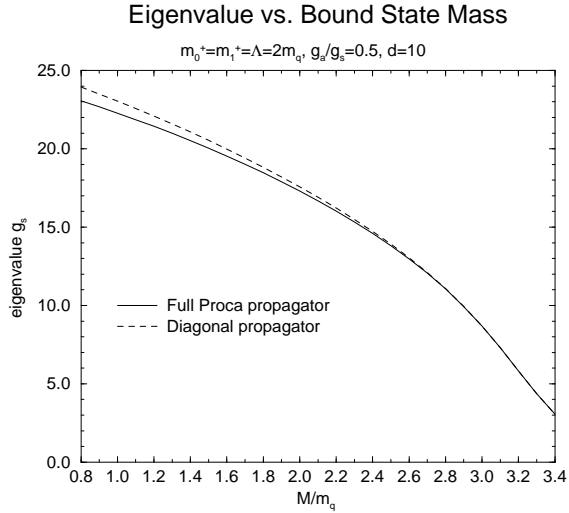}}
\caption{\sf Eigenvalues vs. bound state mass $M$ in the octet equation.
Axialvector diquark propagator with and without diagonal approximation.
\label{procags}}
\end{figure}

In figure \ref{procags} 
we compare the eigenvalues as a function of $M$ for both the 
Proca propagator and its diagonal approximation. Even in regions of moderate 
binding the eigenvalues do not differ by more than 2\% which makes
the approximation of the diagonal propagator in computing the masses a 
reliable one.
While the vertex functions are essentially the same for both
choices of the axialvector diquark propagator, the $A_1$ components
of the Bethe-Salpeter wave function differ by approximately a factor
of 10.  

\section{Direct vs. Modified Ladder Approximation} \label{direct}
\begin{figure}[t]
\centerline{\epsfxsize 7.5cm
\epsfbox{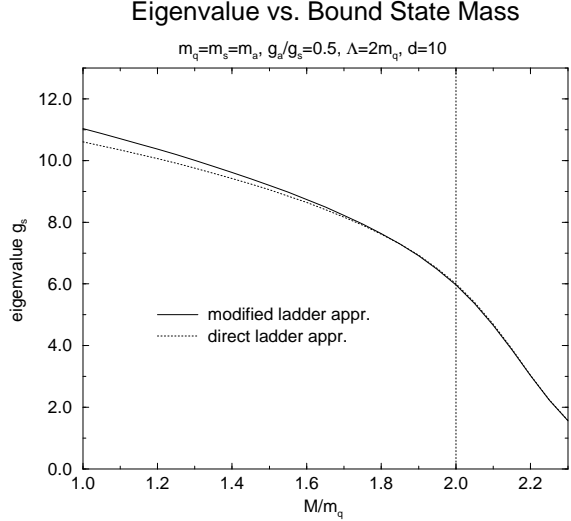}}
\caption{\sf Eigenvalues vs. bound state mass $M$ for direct and
modified ladder approximation.
\label{evdir}}
\end{figure}

In figure \ref{evdir} we display for a representative parameter
set the eigenvalues obtained in direct and modified ladder approximation.
One clearly sees that for bound state masses approximately equal to
the sum of 
the constituent masses the eigenvalues are almost identical, and
even for strongly bound states the deviation is small.
   
\begin{figure}[h]
\centerline{\epsfxsize 5.9cm
\epsfbox{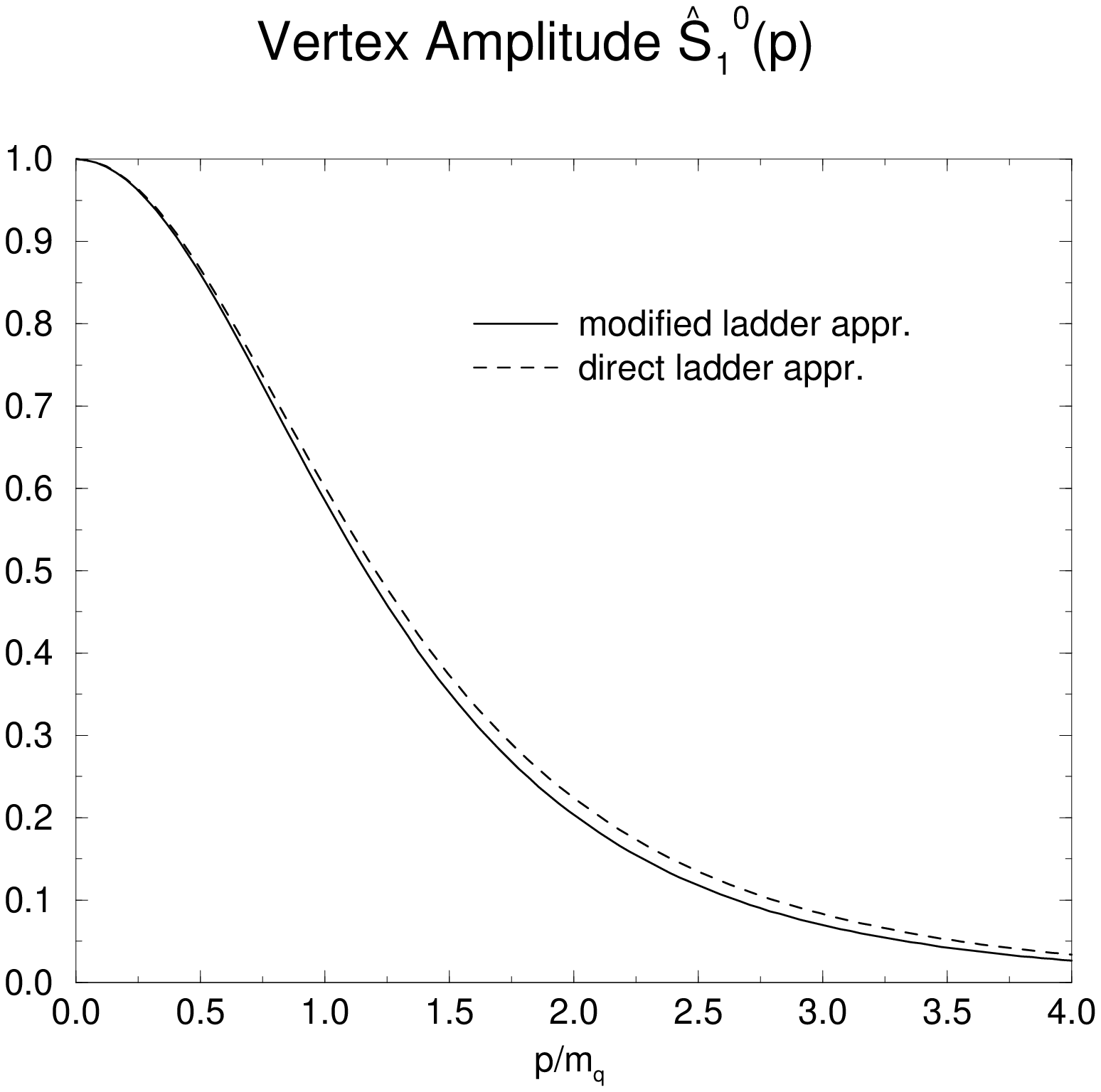}
\epsfxsize 5.9cm
\epsfbox{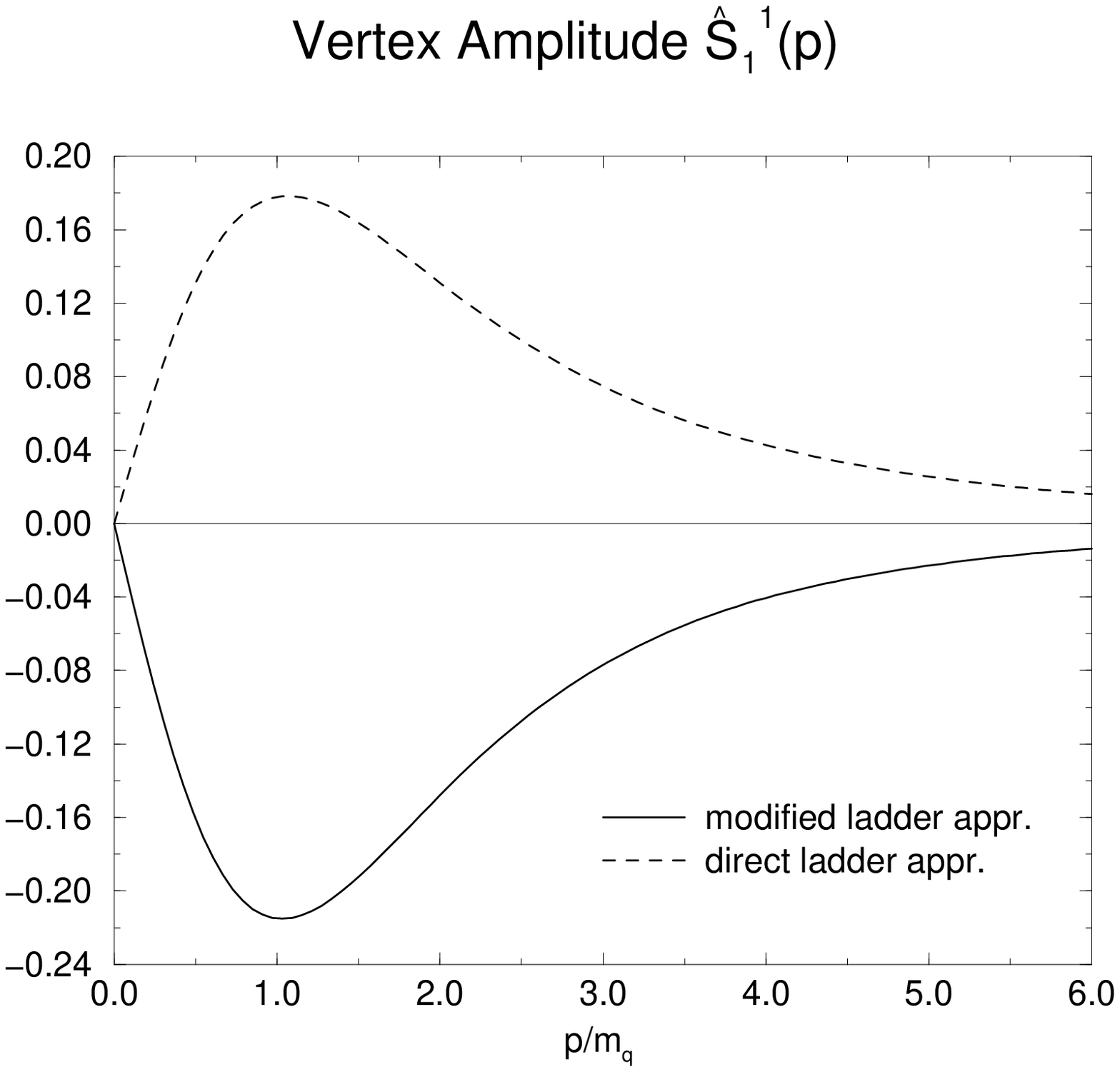}
}
\caption{\sf Zeroth (left) and first (right) Chebyshev moment
of the vertex amplitude $\hat S_1$. Parameters are
$m_q$=$m_{0^+}$=$m_{1^+}$, $M$=1.9$m_q$, $\Lambda$=2$m_q$, $g_a/g_s$=0.5
and $d$=10. \label{sdir}}
\end{figure}

However, larger deviations occur in the vertex functions. Figure
\ref{sdir} shows zeroth and first Chebyshev moment of $\hat S_1$
for modified and direct ladder approximation. Whereas $\hat S_1^0$
hardly differs for the two approaches, $\hat S_1^1$ receives a sign
flip when switching between the two approaches. This occurs for almost
all amplitudes in odd Chebyshev moments. 
When considering the Bethe-Salpeter equation for nucleons with $g_a$=0, the different sign
in $\hat S_1^1$ causes the electric form factor
of the neutron to be changed. Isospin breaking effects are also different:
The neutron-proton mass difference is in the direct approach bigger
than the constituent quark mass difference $m_d-m_u$ whereas in the 
modified ladder approximation it assumes values of approximately
0.7$(m_d-m_u)$ \cite{Oet97}.
\newpage
%
%

%
\end{document}